\pdfoutput=1
\documentclass[12pt,a4paper]{article}

\usepackage{ifthen} 
\newboolean{pdflatex}
\setboolean{pdflatex}{true} 
\newboolean{articletitles}
\setboolean{articletitles}{true} 
\newboolean{uprightparticles}
\setboolean{uprightparticles}{false} 

\def\paperauthors{LHCb collaboration} 
\def\paperasciititle{Amplitude analysis of the radiative decay Bs -> K+K-gamma$} 
\def\papertitle{Amplitude analysis of the radiative decay $B^0_s\to K^+K^-\gamma$} 
\def\paperkeywords{{High Energy Physics}, {LHCb}} 
\def\papercopyright{\the\year\ CERN for the benefit of the LHCb collaboration} 
\def\paperlicence{CC BY 4.0 licence}
\def\paperlicenceurl{https://creativecommons.org/licenses/by/4.0/}

\usepackage[top=1in, bottom=1.25in, left=1in, right=1in]{geometry}

\usepackage{microtype}
\usepackage{lineno}  
\usepackage{xspace} 
\usepackage{caption} 

\usepackage{graphicx}  
\usepackage{color}
\usepackage{colortbl}
\graphicspath{{./figs/}} 

\usepackage{amsmath} 
\usepackage{amssymb}
\usepackage{amsfonts}
\usepackage{upgreek} 

\newcommand*\patchAmsMathEnvironmentForLineno[1]{%
\expandafter\let\csname old#1\expandafter\endcsname\csname #1\endcsname
\expandafter\let\csname oldend#1\expandafter\endcsname\csname
end#1\endcsname
 \renewenvironment{#1}%
   {\linenomath\csname old#1\endcsname}%
   {\csname oldend#1\endcsname\endlinenomath}%
}
\newcommand*\patchBothAmsMathEnvironmentsForLineno[1]{%
  \patchAmsMathEnvironmentForLineno{#1}%
  \patchAmsMathEnvironmentForLineno{#1*}%
}
\AtBeginDocument{%
\patchBothAmsMathEnvironmentsForLineno{equation}%
\patchBothAmsMathEnvironmentsForLineno{align}%
\patchBothAmsMathEnvironmentsForLineno{flalign}%
\patchBothAmsMathEnvironmentsForLineno{alignat}%
\patchBothAmsMathEnvironmentsForLineno{gather}%
\patchBothAmsMathEnvironmentsForLineno{multline}%
\patchBothAmsMathEnvironmentsForLineno{eqnarray}%
}

\usepackage{hyperxmp}

\usepackage[pdftex,
            pdfauthor={\paperauthors},
            pdftitle={\paperasciititle},
            pdfkeywords={\paperkeywords},
            pdfcopyright={Copyright (C) \papercopyright},
            pdflicenseurl={\paperlicenceurl}]{hyperref}

\usepackage[colorinlistoftodos,textsize=scriptsize]{todonotes}

\usepackage[bottom,flushmargin,hang,multiple]{footmisc}

\usepackage[all]{hypcap} 

\usepackage{xspace} 
\usepackage{upgreek}

\def\MagUp {\mbox{\em Mag\kern -0.05em Up}\xspace}

\ifthenelse{\boolean{uprightparticles}}%
{
 
 \def\Pgamma      {\ensuremath{\upgamma}\xspace}

 \def\Ppi         {\ensuremath{\uppi}\xspace}

 \def\Ppsi        {\ensuremath{\uppsi}\xspace}

 \def\PDelta      {\ensuremath{\Delta}\xspace}                 
 \def\PXi         {\ensuremath{\Xi}\xspace}                 
 \def\PLambda     {\ensuremath{\Lambda}\xspace}                 
 \def\PSigma      {\ensuremath{\Sigma}\xspace}                 
 \def\POmega      {\ensuremath{\Omega}\xspace}                 
 \def\PUpsilon    {\ensuremath{\Upsilon}\xspace}
 \let\oldPi\Pi
 \def\PPi         {\ensuremath{\oldPi}\xspace}

 \def\PB      {\ensuremath{\mathrm{B}}\xspace}                 
                  
 \def\PD      {\ensuremath{\mathrm{D}}\xspace}

 \def\PJ      {\ensuremath{\mathrm{J}}\xspace}                 
 \def\PK      {\ensuremath{\mathrm{K}}\xspace}

 \def\PP      {\ensuremath{\mathrm{P}}\xspace}

 \def\Pb      {\ensuremath{\mathrm{b}}\xspace}

 \def\Pi      {\ensuremath{\mathrm{i}}\xspace}

 \def\Ps      {\ensuremath{\mathrm{s}}\xspace}

 \def\thebaroffset{0.0em}
}
{
 
 \def\Pgamma      {\ensuremath{\gamma}\xspace}

 \def\Ppi         {\ensuremath{\pi}\xspace}

 \def\Ppsi        {\ensuremath{\psi}\xspace}                 
                  
 \mathchardef\PDelta="7101
 \mathchardef\PXi="7104
 \mathchardef\PLambda="7103
 \mathchardef\PSigma="7106
 \mathchardef\POmega="710A
 \mathchardef\PUpsilon="7107
 \mathchardef\PPi="7105
                  
 \def\PB      {\ensuremath{B}\xspace}                 
                  
 \def\PD      {\ensuremath{D}\xspace}

 \def\PJ      {\ensuremath{J}\xspace}                 
 \def\PK      {\ensuremath{K}\xspace}

 \def\PP      {\ensuremath{P}\xspace}

 \def\Pb      {\ensuremath{b}\xspace}

 \def\Pi      {\ensuremath{i}\xspace}

 \def\Ps      {\ensuremath{s}\xspace}

 \def\thebaroffset{0.18em}
}
\newcommand{\offsetoverline}[2][\thebaroffset]{\kern #1\overline{\kern -#1 #2}}%

\makeatletter
\ifcase \@ptsize \relax
  \newcommand{\miniscule}{\@setfontsize\miniscule{4}{5}}
\or
  \newcommand{\miniscule}{\@setfontsize\miniscule{5}{6}}
\or
  \newcommand{\miniscule}{\@setfontsize\miniscule{5}{6}}
\fi
\makeatother

\DeclareRobustCommand{\optbar}[1]{\shortstack{{\miniscule (\rule[.5ex]{1.25em}{.18mm})}
  \\ [-.7ex] $#1$}}

\def\g      {{\ensuremath{\Pgamma}}\xspace}

\def\squark    {{\ensuremath{\Ps}}\xspace}

\def\bquark    {{\ensuremath{\Pb}}\xspace}

\def\pion   {{\ensuremath{\Ppi}}\xspace}
\def\piz    {{\ensuremath{\pion^0}}\xspace}

\def\kaon    {{\ensuremath{\PK}}\xspace}
\def\Kbar    {{\ensuremath{\offsetoverline{\PK}}}\xspace}

\def\KorKbar {\kern \thebaroffset\optbar{\kern -\thebaroffset \PK}{}\xspace}

\def\Kpm     {{\ensuremath{\kaon^\pm}}\xspace}

\def\Kstarzb {{\ensuremath{\Kbar{}^{*0}}}\xspace}

\def\Dbar    {{\ensuremath{\offsetoverline{\PD}}}\xspace}
\def\D       {{\ensuremath{\PD}}\xspace}

\def\DorDbar {\kern \thebaroffset\optbar{\kern -\thebaroffset \PD}\xspace}

\def\Dzb     {{\ensuremath{\Dbar{}^0}}\xspace}
\def\Dp      {{\ensuremath{\D^+}}\xspace}
\def\Dm      {{\ensuremath{\D^-}}\xspace}

\def\DpDm    {\ensuremath{\Dp {\kern -0.16em \Dm}}\xspace}

\def\B       {{\ensuremath{\PB}}\xspace}
\def\Bbar    {{\ensuremath{\offsetoverline{\PB}}}\xspace}

\def\BorBbar {\kern \thebaroffset\optbar{\kern -\thebaroffset \PB}\xspace}

\def\Bd      {{\ensuremath{\B^0}}\xspace}

\def\BdorBdbar {\kern \thebaroffset\optbar{\kern -\thebaroffset \Bd}\xspace}

\def\Bs      {{\ensuremath{\B^0_\squark}}\xspace}
\def\Bsb     {{\ensuremath{\Bbar{}^0_\squark}}\xspace}
\def\BsorBsbar {\kern \thebaroffset\optbar{\kern -\thebaroffset \Bs}\xspace}

\def\jpsi     {{\ensuremath{{\PJ\mskip -3mu/\mskip -2mu\Ppsi}}}\xspace}

\def\Y#1S{\ensuremath{\PUpsilon{(#1S)}}\xspace}

\def\Lz          {{\ensuremath{\PLambda}}\xspace}

\def\LorLbar     {\kern \thebaroffset\optbar{\kern -\thebaroffset \PLambda}\xspace}

\def\Lb           {{\ensuremath{\Lz^0_\bquark}}\xspace}

\def\BR         {\BF}

\newcommand{\decay}[2]{\ensuremath{#1\!\to #2}\xspace} 

\def\to                 {\ensuremath{\rightarrow}\xspace}

\def\CP                {{\ensuremath{C\!P}}\xspace}

\def\AT#1     {\ensuremath{A_{\mathrm{T}}^{#1}}\xspace}           

\def\C#1      {\ensuremath{\mathcal{C}_{#1}}\xspace}                       
\def\Cp#1     {\ensuremath{\mathcal{C}_{#1}^{'}}\xspace}                    
\def\Ceff#1   {\ensuremath{\mathcal{C}_{#1}^{\mathrm{(eff)}}}\xspace}        
\def\Cpeff#1  {\ensuremath{\mathcal{C}_{#1}^{'\mathrm{(eff)}}}\xspace}       
\def\Ope#1    {\ensuremath{\mathcal{O}_{#1}}\xspace}                       
\def\Opep#1   {\ensuremath{\mathcal{O}_{#1}^{'}}\xspace}                    

\newcommand{\aunit}[1]{\ensuremath{\text{\,#1}}}       

\newcommand{\tev}{\aunit{Te\kern -0.1em V}\xspace}
\newcommand{\gev}{\aunit{Ge\kern -0.1em V}\xspace}
\newcommand{\mev}{\aunit{Me\kern -0.1em V}\xspace}
\newcommand{\kev}{\aunit{ke\kern -0.1em V}\xspace}
\newcommand{\ev}{\aunit{e\kern -0.1em V}\xspace}
 
\newcommand{\mevc}{\ensuremath{\aunit{Me\kern -0.1em V\!/}c}\xspace}
\newcommand{\gevc}{\ensuremath{\aunit{Ge\kern -0.1em V\!/}c}\xspace}
\newcommand{\mevcc}{\ensuremath{\aunit{Me\kern -0.1em V\!/}c^2}\xspace}
\newcommand{\gevcc}{\ensuremath{\aunit{Ge\kern -0.1em V\!/}c^2}\xspace}

\def\fb   {\ensuremath{\aunit{fb}}\xspace}
\def\invfb   {\ensuremath{\fb^{-1}}\xspace}

\newcommand{\stat}{\aunit{(stat)}\xspace}
\newcommand{\syst}{\aunit{(syst)}\xspace}

\def\gsim{{~\raise.15em\hbox{$>$}\kern-.85em
          \lower.35em\hbox{$\sim$}~}\xspace}
\def\lsim{{~\raise.15em\hbox{$<$}\kern-.85em
          \lower.35em\hbox{$\sim$}~}\xspace}

\def\sPlot{\mbox{\em sPlot}\xspace}

\def\pt         {\ensuremath{p_{\mathrm{T}}}\xspace}

\def\et         {\ensuremath{E_{\mathrm{T}}}\xspace}

\def\tell1  {TELL1\xspace}
\def\ukl1   {UKL1\xspace}

\newcommand{\eg}{\mbox{\itshape e.g.}\xspace}
\newcommand{\ie}{\mbox{\itshape i.e.}\xspace}

\newcommand{\lhcborcid}[1]{\href{https://orcid.org/#1}{\hspace*{0.1em}\raisebox{-0.45ex}{\includegraphics[width=1em]{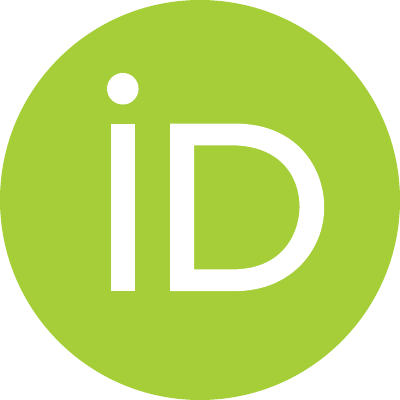}}}}

\usepackage{cite} 
\usepackage{mciteplus}

\newcommand{\ignore}[1]{}

\def\LbpKGam     {\decay{\Lb}{p\kaon^-\g}}
\def\BdKpiGam     {\decay{\Bd}{\kaon^+\pion^-\g}}

\def\BsKKGam     {\decay{\Bs}{\kaon^{+}\kaon^{-} \g}}

\def\fdv{\ensuremath{{\phi{(1020)}}}}
\def\fds{\ensuremath{{f_2{(1270)}}}}
\def\fqv{\ensuremath{{f'_2{(1525)}}}}
\def\fsq{\ensuremath{{\phi{(1680)}}}}
\def\fdc{\ensuremath{{\phi_{\small 3}{(1850)}}}}
\def\fvd{\ensuremath{{f_2{(2010)}}}}
\def\fnr{\ensuremath{{(\textsc{kk})_{\textsc{nr}}}}}

\def\KK      {\ensuremath{K^+K^-}}
\def\KKG     {\ensuremath{K^{+}K^{-}\g~}}

\def\note {\ensuremath{{}^{(*)}}}

\def\diff{\ensuremath{\mathrm{d}}}

\def\gray{\updown} 
\usepackage{xparse}
\usepackage{ifthen}
\usepackage{tikz}
\usepackage{amsmath,lipsum}
\newcommand{\mpm}{\mathbin{\tikz [x=1.2ex,y=1.6ex,line width=.1ex] \draw (0.0,0) -- (1.0,0) (0.5,0.1) -- (0.5,0.9) (0.0,0.5) -- (1.0,0.5);}}%
\NewDocumentCommand\val{mggg}{\ensuremath{ 
    \IfNoValueTF{#4}{}{(}
    \IfNoValueTF{#2}{#1}{\IfNoValueTF{#3}{#1\mpm#2}{\updown {#1~}{}^{\scriptstyle #2}_{\scriptstyle #3}}}
    \IfNoValueTF{#4}{}{)\textrm{#4}}
  }
}
\NewDocumentCommand\valbf{mggg}{\ensuremath{ 
    \IfNoValueTF{#4}{}{(}
    \IfNoValueTF{#2}{\gray\bf #1}{\IfNoValueTF{#3}{\gray\bf #1\mpm#2}{\gray\updown\bf #1~{}^{\scriptstyle\bf #2}_{\scriptstyle\bf #3}}}
    \IfNoValueTF{#4}{}{)\textrm{#4}}
  }
}
\NewDocumentCommand\valb{mggg}{\ensuremath{ 
    \IfNoValueTF{#4}{}{(}
    \IfNoValueTF{#2}{\bf #1}{\IfNoValueTF{#3}{\bf #1\mpm#2}{\updown\bf #1~{}^{\scriptstyle\bf #2}_{\scriptstyle\bf #3}}}
    \IfNoValueTF{#4}{}{)\textrm{#4}}
  }
}

\NewDocumentCommand\rval{mmgg}{\ensuremath{ 
    \IfNoValueTF{#4}{}{(}
    \IfNoValueTF{#3}{\num{#1}\mpm\num{#2}}{\updown \num{#1}~{}^{+\scriptstyle \num{#2}}_{\scriptstyle \num{#3}}}
    \IfNoValueTF{#4}{}{)\textrm{#4}}
  }
}
\NewDocumentCommand\rvalbf{mmgg}{\ensuremath{ 
    \IfNoValueTF{#4}{}{(}
    \IfNoValueTF{#3}{\gray\num[math-rm=\mathbf]{#1}\mpm\num[math-rm=\mathbf]{#2}}{\gray\updown \num[math-rm=\mathbf]{#1}~{}^{+\scriptstyle \num[math-rm=\mathbf]{#2}}_{\scriptstyle \num[math-rm=\mathbf]{#3}}}
    \IfNoValueTF{#4}{}{)\textrm{#4}}
  }
}
\NewDocumentCommand\bias{mmgg}{\ensuremath{ 
    \IfNoValueTF{#4}{}{(}
    \IfNoValueTF{#3}{{\displaystyle [{#1}](\mpm {#2})}}
                {\updown {\displaystyle [{#1}]({}^{\scriptstyle {#2}}_{\scriptstyle {#3}}})}
    \IfNoValueTF{#4}{}{)\textrm{#4}}
  }
}
\NewDocumentCommand\biasbf{mmgg}{\ensuremath{ 
    \IfNoValueTF{#4}{}{(}
    \IfNoValueTF{#3}{\gray\mathbf{\displaystyle [{\bf #1}](\mpm {\bf #2})}}
                {\gray\mathbf\updown {\displaystyle [{\bf #1}]({}^{\scriptstyle {\bf #2}}_{\scriptstyle {\bf #3}})}}
    \IfNoValueTF{#4}{}{)\textrm{#4}}
  }
}

\NewDocumentCommand\rbias{mmgg}{\ensuremath{ 
    \IfNoValueTF{#4}{}{(}
    \IfNoValueTF{#3}{{\displaystyle [\num[retain-explicit-plus]{#1}](\mpm \num{ #2})}}
                {\updown {\displaystyle [\num[retain-explicit-plus]{#1}]({}^{+\scriptstyle \num{#2}}_{\scriptstyle \num{#3}}})}
    \IfNoValueTF{#4}{}{)\textrm{#4}}
  }
}
\NewDocumentCommand\rbiasbf{mmgg}{\ensuremath{ 
    \IfNoValueTF{#4}{}{(}
    \IfNoValueTF{#3}{\gray\mathbf{\displaystyle [\num[retain-explicit-plus,math-rm=\mathbf]{#1}](\mpm \num[math-rm=\mathbf]{ #2})}}
                {\gray\mathbf\updown {\displaystyle [\num[retain-explicit-plus,math-rm=\mathbf]{#1}]({}^{+\scriptstyle \num[math-rm=\mathbf]{#2}}_{\scriptstyle \num[math-rm=\mathbf]{#3}})}}
    \IfNoValueTF{#4}{}{)\textrm{#4}}
  }
}

\NewDocumentCommand\eval{mg}{\ensuremath{ 
    \IfNoValueTF{#2}{ \mpm#1 }{\updown {}^{\scriptstyle #1}_{\scriptstyle #2}}
}}
\NewDocumentCommand\evalbf{mg}{\ensuremath{ 
    \IfNoValueTF{#2}{\gray\bf\mpm #1}{\gray\updown {}^{\bf\scriptstyle #1}_{\bf\scriptstyle #2}}
}}
\NewDocumentCommand\peval{mg}{\ensuremath{ 
    \IfNoValueTF{#2}{\mpm #1}{\updown ({}^{\scriptstyle #1}_{\scriptstyle #2})}
}}

\NewDocumentCommand\erval{mg}{\ensuremath{ 
    \IfNoValueTF{#2}{\mpm \num{#1}}{\updown {}^{+\scriptstyle \num{#1}}_{\scriptstyle \num{#2}}}
}}
\NewDocumentCommand\perval{mg}{\ensuremath{ 
    \IfNoValueTF{#2}{\mpm \num{#1}}{\updown ({}^{+\scriptstyle \num{#1}}_{\scriptstyle \num{#2}})}
}}

\NewDocumentCommand\pervalbf{mg}{\ensuremath{ 
    \IfNoValueTF{#2}{\gray\mathbf\mpm\num[math-rm=\mathbf]{#1}}{\gray\updown ({}^{+\scriptstyle \num[math-rm=\mathbf]{#1}}_{\scriptstyle \num[math-rm=\mathbf]{#2}})}
}}
\NewDocumentCommand\pevalbf{mg}{\ensuremath{ 
    \IfNoValueTF{#2}{\gray\bf \mpm #1}{\gray\updown ({}^{\scriptstyle\bf #1}_{\scriptstyle\bf #2})}
}}

\newcommand{\dw}[2]{\ensuremath{\textrm{d}^{#1}_{#2}}}
\newcommand{\jp}[2]{\ensuremath{#1^{\scriptscriptstyle #2}}}

\def\BR{\ensuremath{{\cal B}}} 
\def\uten{\ensuremath{{\scriptstyle[\times10]}}}

\def\perc{\ensuremath{\%}\xspace}
\newcommand{\kevcc}{\ensuremath{\aunit{ke\kern -0.1em V\!/}c^2}\xspace}
\def\cgev{\ensuremath{\aunit{(Ge\kern -0.1em V\!/}c)^{-1}}\xspace}
\def\rphi{\ensuremath{r_{\phi}}\xspace}
\def\ucgev{\ensuremath{{\scriptstyle\mathrm{[(GeV/c)}^{-1}]}}}
\def\ummev{\ensuremath{{\scriptstyle[\mevcc]}}}

\def\umkev{\ensuremath{{\scriptstyle[\kevcc]}}}
\def\udeg{\ensuremath{{\scriptstyle\rm[deg.]}}}

\def\uperc{\ensuremath{{\scriptstyle[\perc]}}}
\def\muR{\ensuremath{\mu_{\textsc r}}\xspace}
\def\GamR{\ensuremath{\Gamma_{\textsc r}}\xspace}

\def\NLL{\ensuremath{\mathrm{nLL}{}}\xspace}

\def\ibkg{\ensuremath{_{\textsc{bkg}}}}
\def\ik{\ensuremath{_{K}}}
\def\ikk{\ensuremath{_{KK}}}
\def\ikkg{\ensuremath{_{KK\gamma}}}

\def\inr{\ensuremath{_{\textsc{nr}}}\xspace}

\def\mk{\ensuremath{m\ik}\xspace}
\def\mkk{\ensuremath{m\ikk}\xspace}
\def\mkkg{\ensuremath{m\ikkg}\xspace}

\def\tkk{\ensuremath{\theta\ikk}\xspace}
\def\ckk{\ensuremath{\cos\theta\ikk}\xspace}
\def\ackk{\ensuremath{|\cos\theta\ikk|}\xspace}

\def\Bd{\ensuremath{{B^0}}\xspace}

\def\estat{\ensuremath{\mathrm{~(stat.)}}}
\def\esyst{\ensuremath{\mathrm{~(syst.)}}}
\def\ebr{\ensuremath{{~(\mathcal{B})}}}

\def\scr{\ensuremath{\textsc{r}}}
\def\scrp{\ensuremath{{\textsc{r}_\textsc{p}}}}

\def\dr{\ensuremath{{\delta}_\textsc{r}}\xspace}
\def\rFF{\ensuremath{{\cal F}_\textsc{r}/{\cal F}_\phi}\xspace}
\def\FFr{\ensuremath{{\cal F}_\textsc{r}}\xspace}

\def\AA{\ensuremath{{\cal A}}\xspace}
\def\BB{\ensuremath{{\cal B}}\xspace}
\def\FF{\ensuremath{{\cal F}}\xspace}
\def\HH{\ensuremath{{\cal H}}\xspace}
\def\JJ{\ensuremath{{\cal J}}\xspace}
\def\LL{\ensuremath{{\cal L}}\xspace}
\def\NLL{\ensuremath{\mathrm{-ln}{\cal L}}\xspace}
\def\DNLL{\ensuremath{\Delta \mathrm{ln}{\cal L}}\xspace}
\def\MM{\ensuremath{{\cal M}}\xspace}
\def\NN{\ensuremath{{\cal N}}\xspace}

\def\PP{\ensuremath{{\cal P}}\xspace}

\def\VV{\ensuremath{{\cal V}}\xspace}

\def\BW{\ensuremath{{\cal BW}}\xspace}
\def\LLA{\ensuremath{{X}}\xspace}
\def\LLB{\ensuremath{{Y}}\xspace}

\def\sWeight{s\ensuremath{{\cal W}}eight\xspace}
\def\sWeights{s\ensuremath{{\cal W}}eights\xspace}
\def\sWeighted{s\ensuremath{{\cal W}}eighted\xspace}

\def\sPlot{s\ensuremath{{\cal P}}lot\xspace}

\newcommand\up{\rule{0pt}{12pt}}
\newcommand\down{\rule[-6pt]{0pt}{6pt}}
\newcommand\updown{\up\down}
\def\pdf{\textrm{PDF}\xspace}
\def\pdfs{\textrm{PDFs}\xspace}
\def\fit{\textrm{fit}}
\def\simul{\textrm{sim}}
  
\usepackage{longtable} 
\usepackage{booktabs}  
\usepackage{subcaption}
\usepackage{lscape}
\usepackage{float}
\usepackage{graphicx}
\usepackage{titlesec}
\usepackage[title,toc,page]{appendix}
\titleformat{\paragraph}
{\normalfont\normalsize\bfseries}{\theparagraph}{1em}{}
\titlespacing*{\paragraph}
{0pt}{3.25ex plus 1ex minus .2ex}{1.5ex plus .2ex}
\usepackage[version-1-compatibility]{siunitx}
\usepackage{footnote}
\makesavenoteenv{tabular}

\usepackage{longtable} 
\usepackage{enumitem}

\begin{document}

\renewcommand{\thefootnote}{\fnsymbol{footnote}}
\setcounter{footnote}{1}


\begin{titlepage}
\pagenumbering{roman}

\vspace*{-1.5cm}
\centerline{\large EUROPEAN ORGANIZATION FOR NUCLEAR RESEARCH (CERN)}
\vspace*{1.5cm}
\noindent
\begin{tabular*}{\linewidth}{lc@{\extracolsep{\fill}}r@{\extracolsep{0pt}}}
\ifthenelse{\boolean{pdflatex}}
{\vspace*{-1.5cm}\mbox{\!\!\!\includegraphics[width=.14\textwidth]{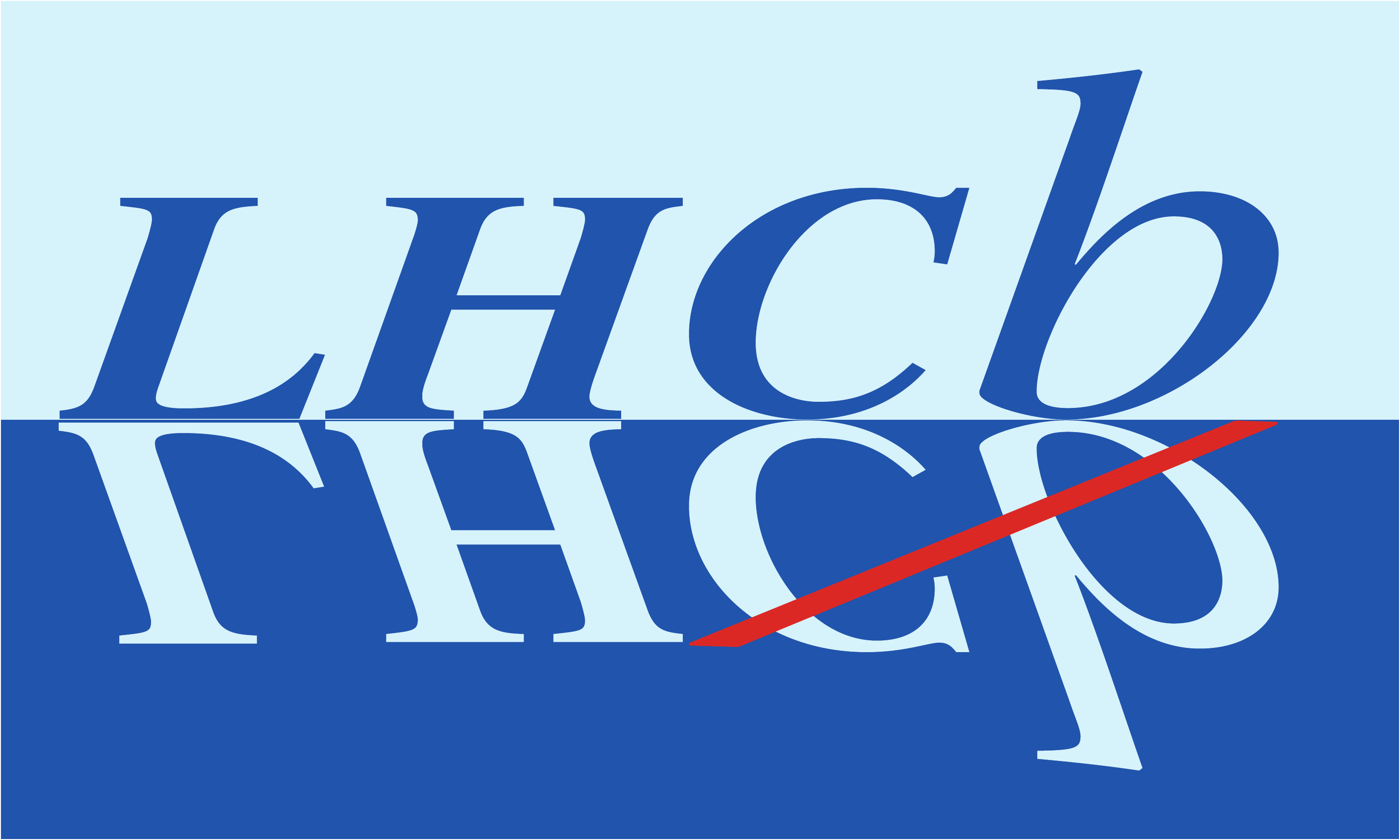}} & &}%
{\vspace*{-1.2cm}\mbox{\!\!\!\includegraphics[width=.12\textwidth]{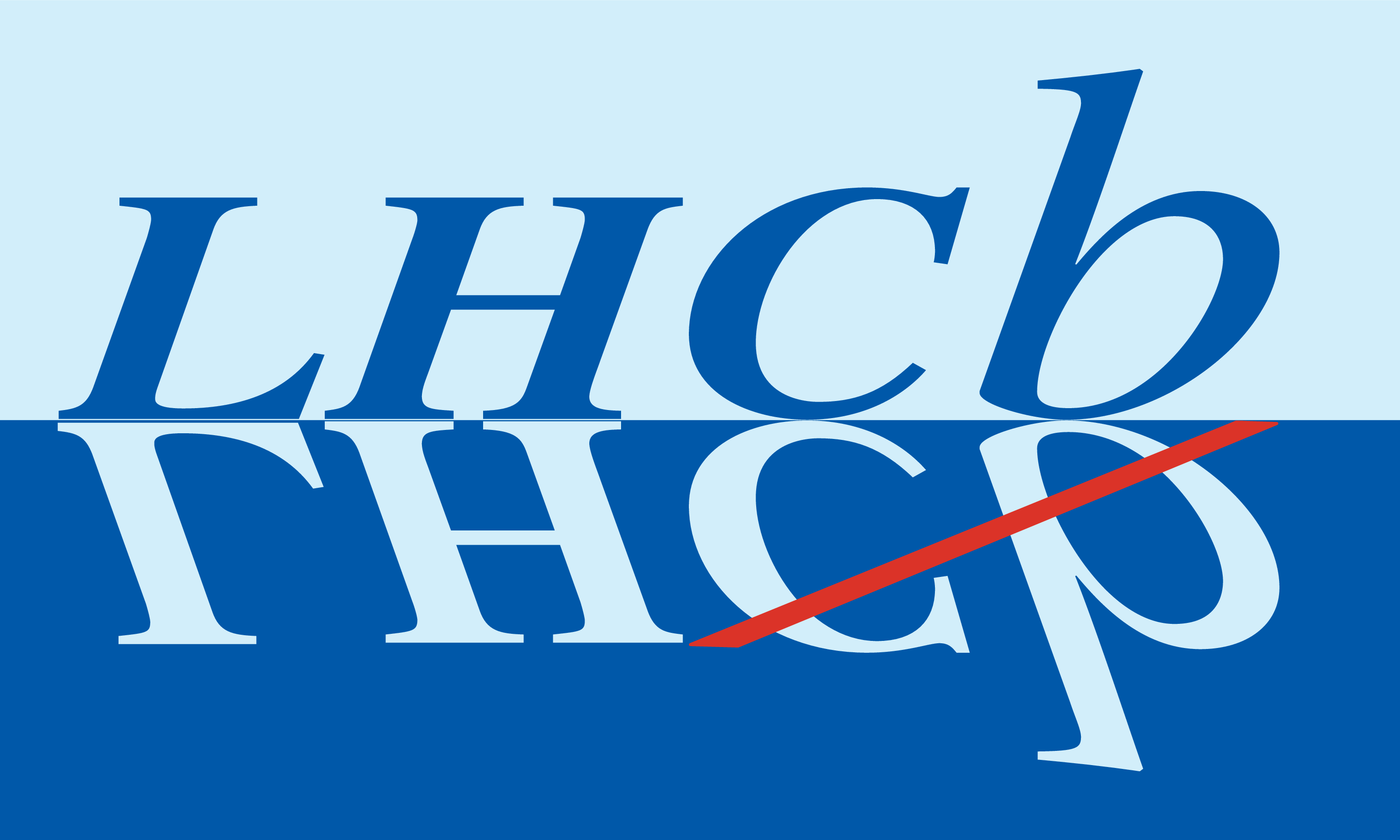}} & &}%
\\
 & & CERN-EP-2024-115 \\  
 & & LHCb-PAPER-2024-002 \\  
 & & August 19, 2024 \\
 & & \\
\end{tabular*}

\vspace*{3.cm}

{\normalfont\bfseries\boldmath\huge
\begin{center}
  \papertitle 
\end{center}
}

\vspace*{1.0cm}

\begin{center}
\paperauthors\footnote{Authors are listed at the end of this paper.}
\end{center}

\vspace{\fill}

\begin{abstract}
  \noindent
  A search for radiative decay of \Bs mesons to orbitally excited $K^+K^-$ states is performed using proton proton collisions recorded by the \mbox{LHCb}\xspace experiment, corresponding to an integrated luminosity of 9\invfb.
  The dikaon spectrum in the mass range $m_{KK}<2400$~{\ensuremath{\,\text{Me\kern -0.1em V\!/}c^2}\xspace} is dominated by the  $\phi(1020)$ resonance that accounts for almost 70$\%$ of the decay rate.
  Considering the possible contributions of $f_2{(1270)}$, $f'_2{(1525)}$ and $f_2{(2010)}$  meson states, the overall tensor contribution to the amplitude is measured to be
\begin{equation}
  {\cal F}_{\{f_2\}}=16.8\mpm0.5\mathrm{~(stat.)}\mpm0.7\mathrm{~(syst.)}\%,\nonumber
\end{equation}
  mostly dominated by the $f'_2(1525)$ state. Several statistically equivalent solutions are obtained for the detailed resonant structure depending on whether the smaller amplitudes interfere destructively or constructively with the dominant amplitude.
  The  preferred solution that corresponds to the lowest values of the fit fractions along with constructive interference leads to the relative branching ratio measurement
\begin{equation}
  \frac{{\cal B}(\Bs\to f'_2\gamma)}{{\cal B}(\Bs\to\phi\gamma)}= 19.4^{+0.9}_{-0.8}\mathrm{~(stat.)}{}^{+1.4}_{-0.5}\mathrm{~(syst.)}\mpm0.5\mathrm{~(\cal{B})}\%\nonumber,
\end{equation}
where the last uncertainty is due to the ratio of measured branching fractions to the $K^+K^-$ final state.
This result represents the first observation of the radiative $\Bs\to f'_2(1525)\gamma$ decay, which is the second radiative transition observed in the $\Bs$ sector. 
\end{abstract}

\vspace*{1.0cm}

\begin{center}
  Submitted to JHEP 08 (2024) 093
\end{center}

\vspace{\fill}

{\footnotesize 
\centerline{\copyright~\papercopyright. \href{\paperlicenceurl}{\paperlicence}.}}
\vspace*{2mm}

\end{titlepage}


\setcounter{page}{2}
\mbox{~}
%
%
%
%


\renewcommand{\thefootnote}{\arabic{footnote}}
\setcounter{footnote}{0}

\cleardoublepage


\pagestyle{plain} 
\setcounter{page}{1}
\pagenumbering{arabic}


\section{Introduction}\label{sec:Introduction}
In the Standard Model (SM), the radiative decays of beauty hadrons proceed at leading order through $b\to s\gamma$ one-loop electromagnetic-penguin transitions, dominated by a virtual intermediate top quark coupled to a $W$ boson. Extensions of the SM predict additional one-loop contributions that can introduce sizeable effects on the dynamics of the transition. Radiative decays of the neutral and charged \B mesons were first observed by the CLEO collaboration in 1993\cite{Ammar:1993sh} through the decay modes $B\to K^{*}\gamma$. In 2007 the Belle collaboration reported the first observation of the companion decay in the $\Bs$ sector\cite{Belle:2014sac}, $\Bs\to\phi\gamma$. The LHC era has brought observations of new radiative $\bquark$-hadron decay modes and precise measurements of branching fractions, helicity structure and asymmetries in this class of decays\cite{LHCb-PAPER-2012-019,LHCb-PAPER-2016-034,LHCb-PAPER-2019-015,LHCb-PAPER-2014-001,LHCb-PAPER-2019-010,LHCb-PAPER-2021-030,LHCb-PAPER-2021-017,LHCb:2024vtc}.

Several exclusive modes have been observed in radiative decays of neutral \Bd mesons\cite{PDG2022}, including tensor intermediate states, and compared to theoretical predictions\cite{Ali_1993,Atwood_1997,VESELI1996309,Ebert:2001en}. In the \Bs sector, the $\phi\gamma$ final state remains the only $b\to s\gamma$ transition observed. Radiative decays of scalar beauty mesons allow a clean spectroscopic representation of the hadronic system accompanying the photon, free of the S-wave amplitude contributions that usually complicate partial wave analyses.\footnote{A similar feature is realised in the $\jpsi\to \KK\pi^0$ hadronic decay\cite{PhysRevD.95.072007}.} This work represents the first amplitude analysis of the dikaon resonant structure in the \BsKKGam decay\footnote{The inclusion of charge-conjugate processes is implicit unless stated otherwise.} up to a dikaon invariant mass ${\mkk=2400\mevcc}$. This analysis exploits  data collected by the LHCb experiment in proton--proton ($pp$) collisions at 7, 8 and 13 TeV centre-of-mass energies in the years of 2011\,--\,2012 (Run 1) and of 2015\,--\,2018 (Run 2), corresponding to 3 \invfb and 6 \invfb of integrated luminosity, respectively.

\section{Detector and selection}\label{sec:Selection}

The LHCb detector is a single-arm forward spectrometer covering the pseudorapidity range $2<\eta<5$  designed for the study of heavy hadrons containing $b$ or $c$ quarks\cite{LHCb-DP-2008-001,LHCb-DP-2014-002}.
The detector elements that are relevant for this analysis are: a silicon-strip vertex detector surrounding the $pp$ interaction region that allows the 
beauty hadron to be identified from its characteristically long flight distance; a tracking system that provides a precise measurement of the dikaon momentum; two ring-imaging Cherenkov
detectors (RICH) that allow to discriminate between the different species of charged hadrons;  a calorimeter system consisting of scintillating-pad (SPD) and preshower detectors, an electromagnetic calorimeter (ECAL) and a hadronic calorimeter, that provides the reconstruction and the identification of the radiated photons. In addition, a muon system allows the identification of muons. 

Simulated samples are used to optimise the selection criteria and evaluate the background contamination. The simulated $pp$ collisions
are generated using Pythia\cite{Sjostrand:2007gs,LHCb-PROC-2010-056_OLD}.
The decay chain of hadronic particles and the final-state radiation are handled by EvtGen\cite{Lange:2001uf} and  PHOTOS\cite{Golonka:2005pn}, respectively.
The detector response to the interacting particles is implemented in the Geant4 framework\cite{Agostinelli:2002hh,Allison:2006ve}. 

The online event selection is performed by a trigger\cite{LHCb-DP-2012-004,trigger_run2}, consisting of a hardware stage based on the information from the calorimeter and muon systems,
followed by a software stage which fully reconstructs the event. In order to reduce  the large level of combinatorial background coming from $pp$ collisions, the hardware trigger selects events having an ECAL
cluster with an energy component transverse to the beam (\et) above a threshold varying between 2.50 and 2.96 (2.11 and 2.70) GeV in Run 1 (Run 2).
To facilitate the reconstruction in the software trigger  the hardware trigger selects only events with fewer than 600 (450) hits in the SPD for Run 1 (Run 2). 

The software trigger is designed to efficiently select candidates with two high transverse momentum (\pt) tracks significantly displaced from the interaction point  and one high-\et photon\cite{topoHLT2}. The trigger efficiency is further enhanced by about $20\%$ by imposing looser track requirements for the events passing a tighter photon threshold, $\et>4$ GeV, at the hardware stage. 
For Run~2 data, a multivariate classifier based on topological criteria complements the software trigger selection\cite{LHCb-PROC-2015-018}.
The Run 1 software trigger requires \mkk to be below 2000 \mevcc. This restrictive criterion has been removed in Run 2 and the dikaon phase space of this analysis is extended up to $\mkk<2400$ \mevcc,  corresponding to the observed phase space of the $\BsKKGam$ signal. A fiducial cut $\mkk<1950$\mevcc, just below the trigger threshold, is applied to the Run 1 data. 

The reconstructed \BsKKGam candidate combines a pair of good-quality tracks and an energetic photon. The two tracks are required to have large impact
parameters (IP) with a significance that exceeds four units with respect to any primary proton--proton collision vertex (PV)\cite{Kucharczyk:1756296}. Both tracks must have a transverse momentum larger than 500\mevc
with at least one above 1.2 \gevc. Kaons are identified using particle identification information provided mainly by the RICH system.
The  probability associated to the kaon hypothesis must be larger than any other hadron hypothesis, pion or proton, and larger than a threshold optimised  to reduce the expected contamination from $\BdKpiGam$ and $\LbpKGam$ radiative decays, which have the same topology as the signal. The optimisation is performed for each year of data taking using simulated samples with particle identification performance derived from dedicated calibration data. The fiducial ranges used for track momentum, $p\in[4.5,100.0]$ \gevc, and pseudorapidity, $\eta\in[1.5,4.5]$, match the phase space covered by the data-driven calibration tool\cite{LHCb-PUB-2016-021}. The two tracks should have a distance of closest approach less than 0.15 mm and form a good quality vertex. Vertex isolation is used to reduce partially reconstructed $\B\to \KK(X)\gamma$ backgrounds, where $X$ generically represents an unreconstructed fragment of the decay final state. Specifically, a lower limit is applied on the $\chi^2$ increase in the  vertex fit when adding any additional reconstructed track, referred to in the following as $\Delta\chi^2_\textrm{Vtx}(\Bs)$. 

Clusters in the ECAL system identified as photon candidates are selected by requiring that they cannot be geometrically associated with any extrapolated track. Photons and neutral pions are distinguished by exploiting
their cluster shape and energy distribution\cite{note-gammapi0calib}. The photon four-momentum is evaluated using the dikaon vertex as the origin and the position and energy of the associated cluster. 
The transverse component of the reconstructed photon momentum is required to  be larger than 3.0\gevc. 

The $\Bs$ candidate four-momentum is computed by summing the four-momenta of the two kaons and the photon. The \Bs candidates are selected in the mass range \mbox{[4700, 6400]\mevcc}.
The momentum is required to point back to the associated primary vertex and to have a transverse component larger than 2.0\gevc. A  significant contamination is expected from the \decay{B^0_{(s)}}{(K^\pm\pi^0)K^{\mp}} decays that involve a $K^\pm\pi^0$ resonant state, including charmed modes through $D^\pm_{(s)}$ decays.
Those hadronic contributions are vetoed by requiring the $\Kpm\gamma$ system to be above the $D_s^\pm$ mass, \ie $m_{K^{\pm}\gamma}>2000\mevcc$,
assigning the neutral pion mass to the reconstructed photon. This criterion, hereafter referred to as the {anti-charm veto}, suppresses the mass peaking contamination from charmless $\Bd\rightarrow K^{*\pm}K^{\mp}$ decays and significantly reduces the partially reconstructed decays involving a misidentified \piz meson.  

Background candidates resulting from combinations of unrelated kaons and photon, hereafter denoted as combinatorial background, can be strongly suppressed by exploiting  kinematic and topological variables.
Boosted Decision Tree classifiers (BDT)\cite{BDT,Roe:2004na}  are trained for each year of data taking using simulated events reproducing the detector conditions as signal proxy and data selected in the upper mass sideband of the signal mass peak as background proxy.
The typical mass resolution of the \BsKKGam signal is 85\mevcc, and is dominated by the photon energy resolution. The upper mass sideband is accordingly defined as $m_{\KK\gamma}> m_{\Bs} + 300$\mevcc, where $m_{\Bs}$ is the known \Bs mass value\cite{PDG2022}. The input variables to the classifier are: the momentum, pseudorapidity, flight distance and $\Delta\chi^2_\textrm{Vtx}(\Bs)$ of the reconstructed \Bs candidate, the IP and transverse momentum of the kaon candidates, the IP, momentum, and \pt of the dikaon combination, and the difference of the primary vertex fit $\chi^2$ calculated with or without the tracks associated to the reconstructed $\Bs$ meson. An additional input variable to the BDT in Run 2 is the isolation variable,
\begin{equation}
  I_{\pt}=\frac{\pt(\Bs)-\sum \pt}{\pt(\Bs)+\sum \pt},
\end{equation}
where the sum is taken over tracks that are not part of the $\Bs$ signal candidate but are associated with the same PV and fall within a cone of half-angle $\Delta R<$ 1.7 rad.
The half-angle of a track is defined as $(\Delta R)^2 = (\Delta\theta)^2 +(\Delta\phi)^2$, where $\Delta\theta$ and $\Delta\phi$ are the differences in the polar and
azimuthal angles of each track with respect to the $\Bs$ candidate direction. The optimal BDT working point is optimised for each year of data taking by maximising the ratio $S/\sqrt{S+B}$,
where $S$ is the  expected number of signal candidates estimated from simulation and $B$ is the number of combinatorial background candidates in the signal region
estimated by extrapolating the data distribution in the upper mass sideband of the signal mass peak. The efficiency of the  optimal BDT cut on the preselected signal is around 97\%
while the combinatorial background is reduced by factor of 20. 

The remaining combinatorial background  and the partially reconstructed $B$ decays can be constrained by their invariant-mass distribution on both sides of the signal peak.
The selected $\KKG$ sample is additionally polluted by misidentified $\BdKpiGam$ and  $\LbpKGam$ decays that pass the dikaon identification requirements and populate the \Bs signal region. The corresponding $\Bd$ and $\Lb$ contamination estimated from simulation are \val{4.5}{1.1\perc} and \val{6.9}{1.9\perc} of the signal yield, respectively. Due to the limited calorimeter energy resolution and the resulting wide signal peak, the misidentified  backgrounds cannot be efficiently separated from the \Bs signal mass distribution.  Further peaking backgrounds stemming from photon misidentification, such as $\B\to \KK\pi^0$ charmless decays, are highly suppressed thanks to the {anti-charm veto} that rejects the $K^{*\pm}K^\mp$ intermediate states. The residual colour-suppressed decay modes to the $(\KK)\pi^0$ final states are difficult to quantify due to their unknown resonant structure. The suppressed charmed decay $B^0\to \Dzb\pi^0$ with $\Dzb\to \KK$ or $K^+\pi^-$, which is well-localized in the dikaon mass spectrum, can similarly contribute to the signal mass region. The unresolved peaking contributions are, therefore, embedded in the signal component in the mass model and their description is handled in the subsequent amplitude analysis of the dikaon system.
 
\section{Invariant-mass fit}\label{sec:MasFit}
The invariant-mass distribution of the \BsKKGam signal is modelled using a modified double-sided Crystal-Ball\cite{osti_5211117} probability density function (\pdf) with an asymmetrical Gaussian core and tails on either side.
The Gaussian mass peak position and the left/right width parameters are allowed to vary freely in the fit to accommodate the possible difference between simulation and data resolutions and to account for the contamination of the embedded peaking backgrounds.
The low-mass tail parameters cannot be resolved in the fit to data due to the large partially reconstructed backgrounds populating the left sideband. They are fixed to the values obtained from a fit to simulated samples. The high-mass tail accounts for the imperfections of the tracking and, in the case of radiative decays, the large cluster pile-up variations in the ECAL which may affect the photon energy determination. The parameter that defines the location of transition to the right-tail function is allowed to vary freely in the fit to adjust the misidentified $\LbpKGam$ background, mostly contaminating the right side of the signal peak, while the tail decay parameter is fixed to the value obtained from simulation. The partially reconstructed $\Bs\to \KK(X)\gamma$ backgrounds are described using ARGUS functions\cite{ARGUS} convoluted with the signal resolution function.
Two inclusive components, with one or with two missing pions, are considered  for the partially reconstructed $\Bs$ decays in the nominal mass range, {$\mkkg\in [4700,6400]$\mevcc}. Several exclusive decay modes with similar shapes potentially compete in the partially reconstructed decay region. Furthermore, the branching fractions of most of these decays are unknown. The overall yield of the inclusive partially reconstructed contributions, hereafter denoted as one-missing-pion and two-missing-pion components, is thus allowed to vary freely in the fit to data. The shape of the one-missing-pion component is partially constrained using the parameterisation obtained from a fit to the simulated decay $\Bs\to\fsq(\to \KK\pi^0)\gamma$, used as a proxy. In the fit to data, the curvature parameter of the ARGUS function is fixed to the value obtained from the fit to this simulated sample. The missing-mass shift is set to the known neutral pion mass\cite{PDG2022}, $m_{\pi^0}$,  and the slope parameter, which depends on the actual decay dynamics, is allowed to vary freely to accommodate the unknown composition of the generic one-missing-pion component. The two-missing-pion component, which mostly contributes to the lower edge of the mass window, is modelled using a similar ARGUS function, with a missing-mass shift fixed to $2m_{\pi^0}$ and a free slope parameter. The curvature parameter, poorly resolved in the fit to data, is fixed to the same value as for the one-missing-pion component. The combinatorial background due to random $KK\gamma$ combinations is modelled using a decreasing exponential shape where the decay parameter is free to vary. The peaking backgrounds are not modelled separately and are included in the signal component, as previously discussed. 

An unbinned extended maximum-likelihood fit is performed according to the \pdf
\begin{equation}
\mathcal{F}(\mkkg)={N}_\textsc{s}\cdot {S}(\mkkg)+\sum_{\textsc{bkg}} {N}_{\textsc{bkg}}\cdot {B}_{\textsc{bkg}}(\mkkg),
\label{eq:fit_func}
\end{equation}
where ${S}$  (${B}_{\textsc{bkg}}$) represents the signal (background) \pdf and ${N}_\textsc{s}$ (${N}_{\textsc{bkg}}$) the associated yield(s) allowed to vary freely.
\begin{figure}[htbp]
\centering
\includegraphics[width=25pc]{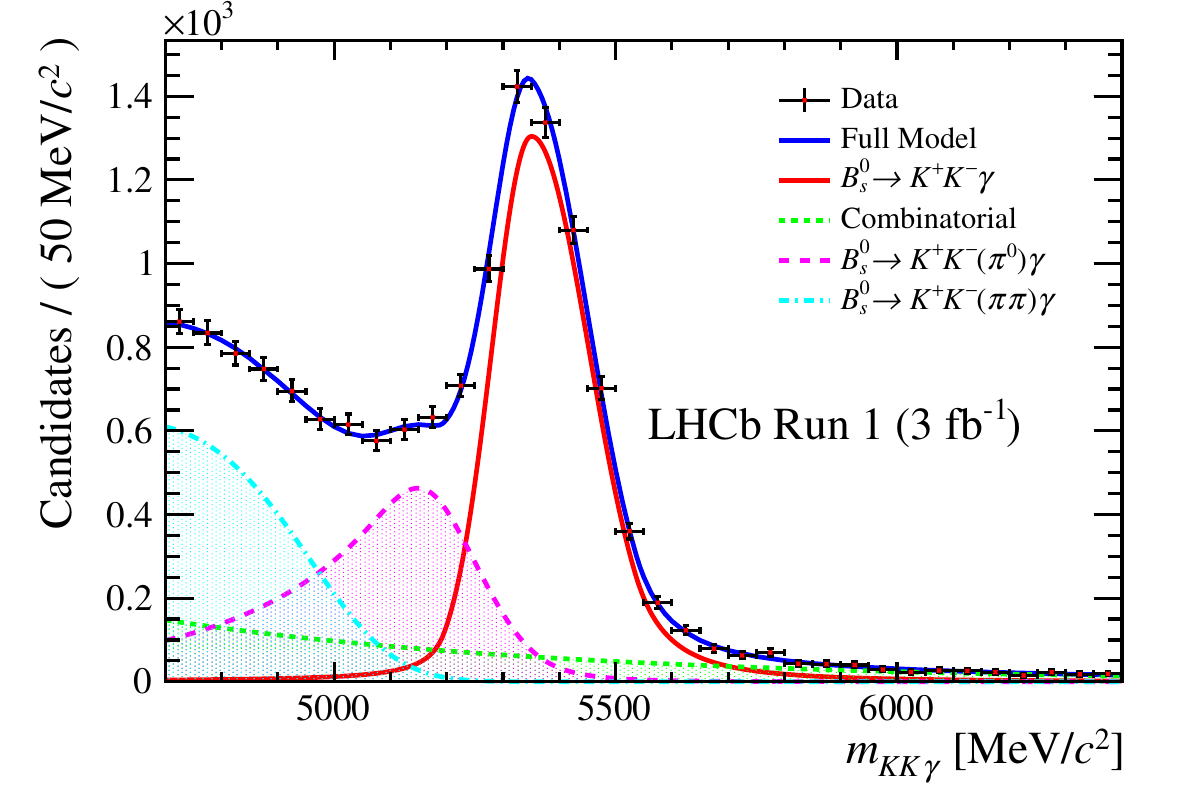}
\includegraphics[width=25pc]{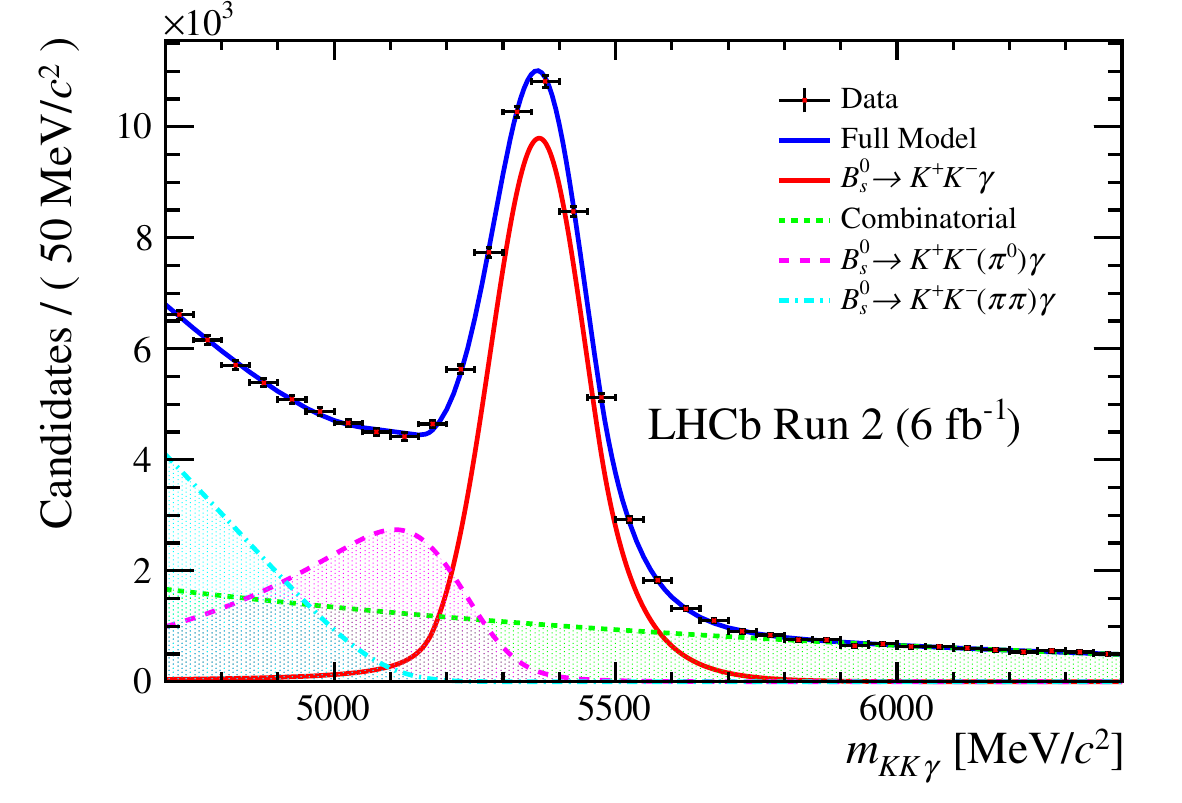}
  \caption{Invariant-mass distribution for $K^{+}K^{-}\gamma$ candidates for (top) Run~1 and (bottom) Run2, with the fit projection overlaid. 
}\label{fig:MasFit}
\end{figure}
The resulting fit projections on the Run 1 and Run 2 data samples are shown in Fig.~\ref{fig:MasFit}. The yield of the $\BsKKGam$ signal candidates is found to be $N_\textsc{s}= (\val{5.66}{0.14})\times 10^3$ and $(\val{44.5}{0.5})\times 10^3$ in Run 1 and Run 2, respectively, including  peaking background components that are expected to contribute about $10\%$. 

\begin{figure}[htbp]
\centering
\includegraphics[width=18.pc]{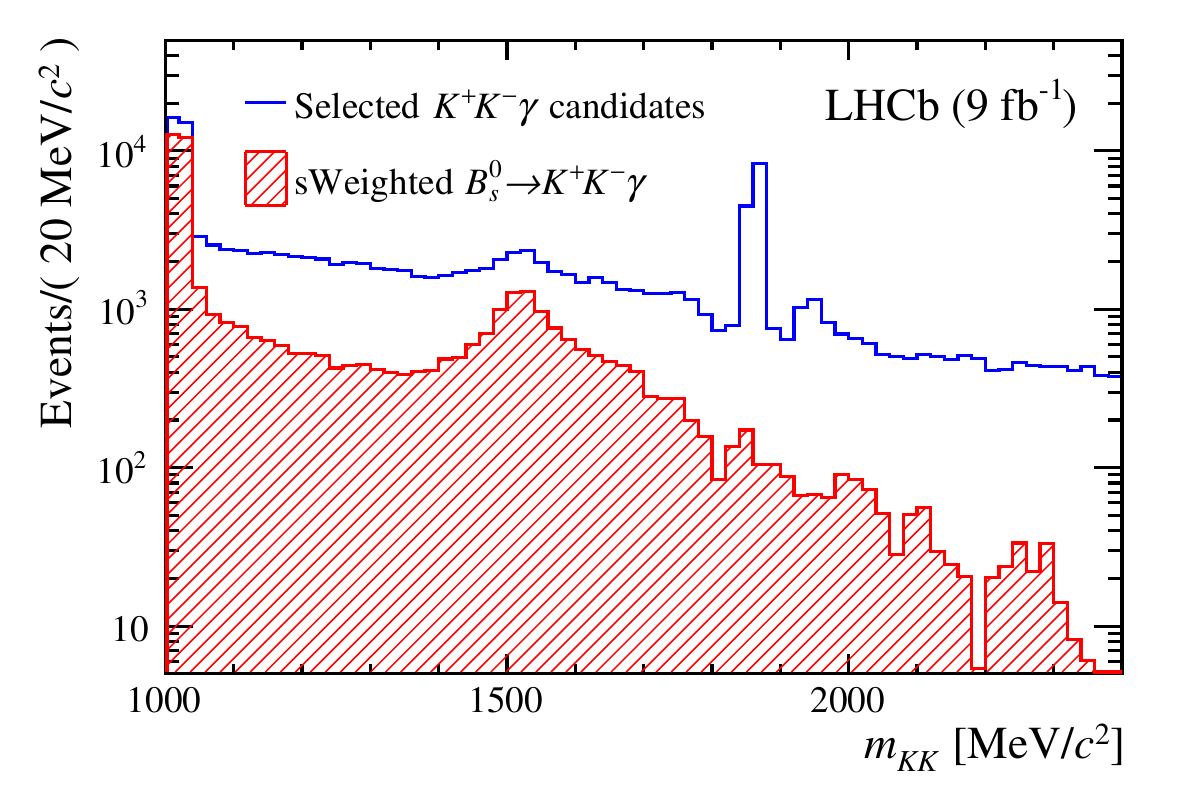}
\includegraphics[width=18.pc]{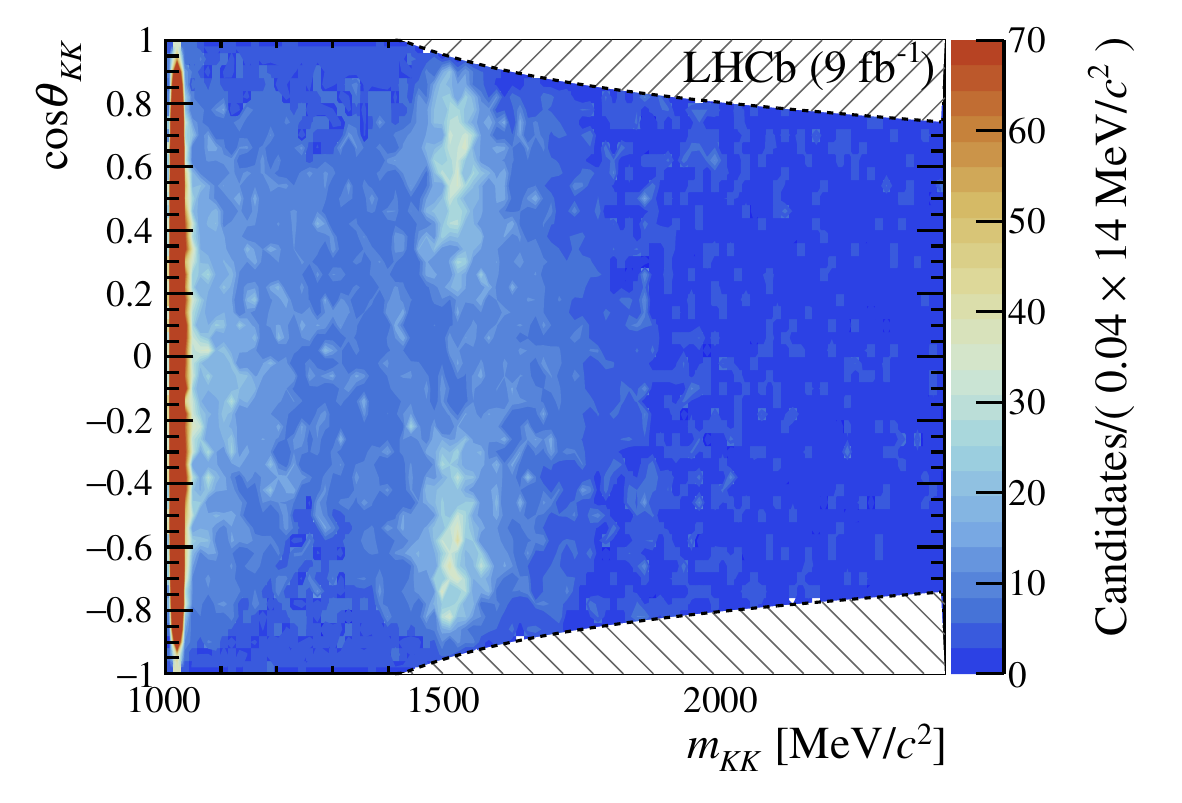}
\caption{(Left) dikaon invariant mass for the selected $K^{+}K^{-}\gamma$ candidates and the s$\cal W$eighted signal distribution. The   narrow peak around 1860 \mevcc and its small reflection 100 \mevcc above correspond to $D^0\to K^+K^-$  and to misidentified $\D^0\to K^-\pi^+$ decays, which are  strongly suppressed in the s$\cal W$eighted distribution.
  (Right) s$\cal W$eighted projection of the ${\Bs\to K^{+}K^{-}\gamma}$ signal on the $(m_{KK},\cos\theta_{KK})$ plane. The hatched areas indicate the acceptance regions suppressed by the anti-charm veto (see Sect.~\ref{ssec:acceptance} for details).}\label{fig:mkk}
\end{figure}

Following the $s\mathcal{P}$lot technique\cite{splot}, a signal weight (\sWeight) is assigned to each candidate to statistically subtract the combinatorial and partially reconstructed background components in the subsequent amplitude analysis. The left-hand plot in Fig.~\ref{fig:mkk} displays the dikaon mass distribution for the selected $\KKG$ candidates.  The \sWeighted projection of the signal component that contributes up to $\mkk\sim 2400$ \mevcc is superimposed.
The right-hand plot in Fig.~\ref{fig:mkk} displays the signal \sWeighted projection on the amplitude observables plane $(\mkk,\ckk)$, where $\tkk$ is the helicity angle defined as the angle between the positively charged kaon direction and the $\Bs$ meson momentum in the dikaon rest frame. Aside from the dominant vector contribution in the $\phi(1020)$ region, one can clearly see a tensor contribution around 1500\mevcc, identifiable as a significant contribution from the $f'_2(1525)$ resonance.

\section{Amplitude analysis}\label{sec:AmpAn}
\subsection{Amplitude model}\label{ssec:AmpModel}
The kinematics of the three-body transition $B^0_s\to(\KK)_\scr\gamma$, where R is an intermediate dikaon state, can be completely described by the invariant dikaon mass, $\mkk$, and the helicity observable, $\ckk$, related to the traditional Dalitz\cite{Dalitz} mass coordinates $s_{ij}=m^2_{ij}$ as
\begin{equation}
\ckk=\frac{(s_{K^+\gamma}-s_{K^-\gamma})\mkk}{4 M_{\Bs} q_\scr q_{\Bs}}c^2\label{DalitzHelicity},
\end{equation}
where $q_\scr$ and $q_\Bs$ represent the kaon momentum in the dikaon rest frame and the dikaon momentum in the $\Bs$ rest frame, respectively.
The Lorentz-invariant three-body decay rate in the $(\mkk,\ckk)$ coordinates system is given by
\begin{equation}
  \diff\Gamma= {\cal J}_3(\mkk)|\MM(\mkk,\ckk)|^{2}\diff\mkk \diff\!\ckk,
\end{equation}
where ${\cal J}_3(\mkk)$ represents the three-body phase-space Jacobian\footnote{Irrelevant constant factors are omitted here.}
\begin{equation}
  {\cal J}_3(\mkk)\propto\frac{q_\scr q_\Bs}{M_\Bs^2}c^{-2},
\end{equation}
and the matrix element $\MM(\mkk,\ckk)$ represents the transition amplitude. The transition probability is obtained by summing incoherently over the unobserved photon helicity states
\begin{equation}
  |\MM|^2=\sum_{\lambda=\pm 1}|\MM_\lambda|^2=2|\MM_{|\lambda|=1}|^2,
\end{equation}
where the last identity results from the fact that $\MM_{\lambda=+1}$ exhibits the same $\tkk$ helicity dependency as $\MM_{\lambda=-1}$. 

To describe this amplitude transition, an isobar approach is used that consists of the coherent sum of the individual amplitudes describing the intermediate states
\begin{equation}
\MM(\mkk,\ckk)=\sum_\scr c_\scr~\AA_\scr(\mkk,\ckk),
\end{equation}
where $c_\scr$ is a complex coefficient and $\AA_\scr$ represents the amplitude for the intermediate state {\sc R}. The amplitudes $\AA_\scr$ are modelled as
\begin{equation}
  \AA_\scr(\mkk,\ckk)= A_\scr(\mkk)\dw{J_\scr}{10}(\ckk),
\end{equation}
where  $A_\scr(\mkk)$ is the mass lineshape of the intermediate state {\sc R} with spin $J_\scr$ .The angular dependency is given by the Wigner d-functions \dw{J_\scr}{\lambda\lambda'}(\tkk) that represent the matrix elements of the operator rotating the angular momentum basis from the $\Bs$ decay axis ($|\lambda|=1$) to the dikaon decay axis ($\lambda'=0$) \cite{Martin_1960}.
As no S-wave is allowed in the radiative decays of \B mesons, a nominal model based on relativistic Breit--Wigner amplitudes is adopted to describe the mass lineshapes, $A_\scr(\mkk)$, for all the considered resonant states
\begin{equation}
  A_\scr(\mkk)=\FF_\scr\cdot\FF_B\cdot\BW_\scr(\mkk;\mu_\scr,\Gamma_\scr),
\end{equation}
where $\FF_\scr$ and $\FF_B$ are the  Blatt--Weisskopf factors\cite{BlattWeisskopf} accounting for the centrifugal barrier effect in the decays of the {\sc R} resonance and the \Bs meson, respectively.
The Breit--Wigner complex pole for the resonance {\sc R} is given by
\begin{equation}
  \BW_\scr(\mkk;\mu_\scr,\Gamma_\scr)=\frac{1}{ (\mu_\scr^2-\mkk^2)-i\mu_\scr{\cal W}(\mkk;\Gamma_\scr)},
\end{equation}
with $\mu_\scr$ and $\Gamma_\scr$, are the corresponding pole mass and width. The mass-dependent width is defined as
\begin{equation}
  {\cal W}(\mkk;\Gamma_\scr)=\Gamma_\scr\frac{q_\scr}{\bar q_\scr}\frac{\mu_\scr}{\mkk}\FF_\scr^2,
\end{equation}
where $\bar q_\scr$ is a reference kaon momentum evaluated at the nominal mass pole of the resonance. The normalized Blatt--Weisskopf form-factors
\begin{eqnarray}
  \FF_\scr&=&\FF(q_\scr,\bar q_\scr,L_\scr),\\
  \FF_{B}&=&\FF(q_{B},\bar q_B,L_{B}),
\end{eqnarray}
are derived from the spherical Hankel functions of first kind\cite{BlattWeisskopf},
\begin{equation}
  \FF(q,\bar q,L)=\left|\frac{\HH_L(r\bar q)}{\HH_L(rq)}\right|=\left( \frac{q}{\bar q}\right)^L \frac{h_L(r\bar q)}{h_L(rq)},
\end{equation}
where the parameter $r$ is the meson radius that accounts for the size of the centrifugal barrier effect and $L$ is the relative angular momentum in the resonance decay.
The $L$-dependent functions $h_L(z)$ for $L\le 4$ are
\begin{eqnarray}
h_0(z)&=&1  ,\\
h_1(z)&=&\sqrt{1+z^2}  ,\\
h_2(z)&=&\sqrt{9+3z^2+z^4}  ,\\
h_3(z)&=&\sqrt{225+45z^2+6z^4+z^6} ,\\
h_4(z)&=&\sqrt{11025+1575z^2+135z^4+10z^6+z^8}.
\end{eqnarray}
The radius parameter associated with the $\fdv$ meson lineshape, $r_\phi$, is allowed to vary in the fit. The radius value is fixed to $r_{f_2}=3.0$~\cgev for heavier dikaon resonances, and to $r_B=5.0$~\cgev for the $\Bs$ meson. The variation of the radii of heavy mesons is considered when evaluating systematic uncertainties.
The relative angular momentum $L_\scr$ of the pseudoscalar kaons equals the resonance spin. The relative angular momentum of the resonance in the radiative $B^0_s$ decay takes eigenvalues in the $J_\scr\pm 1$ range. The lowest allowed value, $J_\scr-1$, is assumed in the nominal model. Other  values are considered when evaluating systematic uncertainties. 

The experimental dikaon mass resolution is included in each individual Breit--Wigner pole using an analytical approach derived from Ref.~\cite{Kycia} and assuming Gaussian behaviour. 
The relativistic Voigt profile is built by convoluting the Breit--Wigner profile with a  Gaussian resolution function 
\begin{equation}
  |\VV(m ; \mu, \Gamma, \sigma)|^2=
  \int_{-\infty}^{+\infty}  |\BW(m;\mu,\Gamma)|^2  {\cal G}(m-m^\prime;0,\sigma)  \diff m^{\prime}=\frac{1}{\sigma \sqrt{2 \pi}}|\HH(a,u_+,u_-)|^2, 
\end{equation}
with $u_\pm=\frac{m\pm\mu}{\sqrt{2}\sigma}$ and $a=\frac{\mu{\cal W}(m;\Gamma)}{2\sigma^2}$, and $\mu$ and $\sigma$ representing the Gaussian parameters.
The result of the integration is a weighted-sum of Faddeeva functions
\begin{eqnarray}
  |\HH(a, u_{+}, u_{-})|^2
  &=&\frac{w(z_{++})+w(z_{+-})}{2 \Delta_{+}}+ \frac{w(z_{-+})+w(z_{--})}{2 \Delta_{-}},
\end{eqnarray}
where $z_{\kappa\eta}=(u_++u_- + \kappa\cdot\Delta_\eta)/2$, $\Delta_\eta=\sqrt{(u_+-u_-)^2+\eta\cdot 4ia}$ and $w(z)$ is the  Faddeeva function, \ie the scaled complementary error complex function, the real part  of which defines the usual nonrelativistic Voigt profile
\begin{equation}
  w(z)=e^{-z^2} \operatorname{erfc}(-i z). 
\end{equation}
The mass resolution is then included in the amplitude model by redefining the Breit--Wigner pole definition as
\begin{equation}
\BW_R(\mkk;\mu_\scr,\Gamma_\scr,\sigma_\scr)=|\VV(\mkk;\mu_\scr,\Gamma_\scr,\sigma_\scr)|e^{i\rm{Arg}\left[\BB_R(\mkk;\mu_\scr,\Gamma_\scr)\right]},
\end{equation}
\ie the resolution is included in the mass lineshape, but the effect of the resolution on the mass-dependent phase is neglected. The nominal resolution values derived from simulation studies are fixed to $\sigma_\phi=0.54$ \mevcc for the $\phi(1020)$ lineshape, and to $\sigma_\scr=3.2$ \mevcc for all the higher-mass resonances. The experimental resolution on the helicity observable $\tkk$, found to be negligible over the whole analysis range, is not introduced in the model. 

As the $\KK$ system is a \CP eigenstate, the flavour of the decaying $B^0_s$ meson is undefined in this time-integrated analysis. The helicity angle observable in the symmetrical $\KK$ system, $\theta_H$, is univocally defined as the angle between the momentum of the  positively-charged kaon and the \Bs  momentum in the dikaon rest frame. This measured angle matches the helicity angle  for one of the $\Bs$ flavours,  $\theta_H=\tkk$, but corresponds to the opposite angle, $\theta_H=\pi-\tkk$, for the opposite flavour. As a consequence, the interference between odd- and even-spin components, which is an anti-symmetrical function of the helicity, cancels out in the case of equal decay rates of the two flavours. Thanks to the fast $\Bs$ oscillation, any small flavour asymmetry at the production level\cite{LHCb-PAPER-2014-042} is  diluted to a negligible level when integrating over time. Assuming, in addition, that there is no violation of the \CP symmetry in the penguin-mediated radiative decay~\cite{physrevd.72.014013}, an equal decay rate for \Bs and \Bsb is expected. Residual experimental asymmetries due, for instance, to differences in the momentum-dependent $K^+/K^-$ detection efficiencies are explicitly cancelled out by considering the folded (\mkk,$|\ckk|$) half plane and summing incoherently the odd- and even-spin amplitude subsystems. The nominal probability density function  describing the signal component is then defined as
\begin{equation}
  \PP_s(\mkk,\tkk)=\varepsilon(\mkk,\tkk)\cdot\JJ_3(\mkk)\sum_{\textsc{p}=+,-}\left|\sum_{\scrp} c_{\scrp} \cdot \AA_{\scrp}(\mkk,|\ckk|)\right|^2,\label{eq:sigpdf}
\end{equation}
where $\AA_{\scrp}$ is the amplitude for the component $\textsc{R}_\textsc{p}$ with spin parity $\textsc p$,  $c_{\scrp}=|c_{\scrp}|e^{i\delta_{\scrp}}$ is the associated complex isobar coefficient and $\varepsilon(\mkk,|\ckk|)$ is the parametrised experimental acceptance presented in Sect.~\ref{ssec:acceptance}. 

Including backgrounds, the full \pdf describing the selected data sample is given by
\begin{equation}
  \PP(\mkk,\tkk)=\NN_\textsc{s}\cdot \PP_\textsc{s}+\sum\ibkg\NN\ibkg\cdot\PP\ibkg, \label{eq:pdf}
\end{equation}
where $\NN_\textsc{s}$ is the overall $(\Bs+\Bsb)\to \KKG$ yield, and $\NN\ibkg$ and $\PP\ibkg$ represent the yields and PDFs of the backgrounds, respectively. Each \pdf component, $\PP_\text{s}$ and $\PP\ibkg$, is normalised to unity.

\subsection{Acceptance}\label{ssec:acceptance}
 The two-dimensional selection acceptance $\varepsilon(\mkk, \ackk)$ is determined from a simultaneous fit to large samples of fully reconstructed \BsKKGam simulated decays, uniformly produced in the decay phase space, $\Bs\to \fdv\gamma$ decays and $\Bs\to\fqv\gamma$  decays.
Weights are applied to the simulated candidates to correct for  imperfections in the simulation of kinematic variables and to reproduce the neutral and charged particle identification efficiencies using a data-driven calibration\cite{LHCb-PUB-2016-021,note-gammapi0calib}. 

The \pdf that describes each of the simulated samples,
\begin{equation}
  \PP_\scr=\varepsilon(\mkk,\tkk;{\vec\alpha})\cdot \JJ_3(\mkk)\cdot |A_\scr(\mkk)\dw{J_\scr}{\lambda,0}(\tkk)|^2 ,
\end{equation}
embeds the parameterised two-dimensional acceptance function, $\varepsilon(\mkk,\tkk;{\vec\alpha})$. A generic acceptance function based on Bernstein polynomials up to degree 5 is defined as
\begin{equation}
  {\varepsilon}(\textsc x,\textsc y;\vec{\alpha})=1+\left[\sum_{i=0}^4\sum_{j=0}^5\alpha_{ij}\cdot{ B}_i^4(\textsc y)\cdot{B}_j^5(\textsc x)\right]
    \cdot{\varepsilon}_b(\mkk,|\ckk|;\sigma_c),
  \label{Eq:AccFun}
\end{equation}
where the coordinate $\textsc x={q_\scr(\mkk)/}{q_0}$ is the normalized kaon momentum in the dikaon rest frame and $\textsc y=|\ckk|$ is the folded helicity observable. 
The functions ${B}_k^{n}(u)$ represent the ${n}^{th}$-order Bernstein polynomials
\begin{equation}
{B}_k^{n}(u)= {n \choose k} u^k(1-u)^{n-k}.
\end{equation}
The normalisation  factor $q_0=q_\scr(\mkk^{\textrm max})$ ensures the Bernstein argument $\textsc x$ lies within the $[0,1]$ bounds for the considered mass range, $\mkk\in [2\mk,\mkk^{\textrm max}]$.
Conditions are  applied to the $\alpha_{ij}$ parameters such that the acceptance becomes independent of the undefined helicity value at the decay threshold $\textsc x=0$  (implying $\alpha_{i0}=0,~\forall i$) and  that the acceptance derivative is continuous on the helicity folding line $\textsc y=0$ (implying $\alpha_{0j}=\alpha_{1j},~ \forall j$). These two constraints leave twenty independent $\alpha_{ij}$ parameters to be determined in the fit. 

The second factor appearing in the acceptance function of Eq.~\ref{Eq:AccFun}, is defined as
\begin{equation}
{\varepsilon}_b(\mkk,|\ckk|;\sigma_c)=\frac{1}{2}\left[1+\textrm{erf}\left(\frac{\textsc c_0(\mkk)-|\ckk|}{\sigma_c}\right)\right],
\end{equation}
and aims to describe the effect of the {anti-charm veto} that directly affects the Dalitz acceptance region. This criteria, $m_{K^\pm\gamma\to \pi^0}>m_{\rm cut}$, is equivalent to a mass-dependent helicity range
\begin{equation}
  |\ckk| < \textsc c_0(\mkk)\equiv\frac{\left(M_{\Bs}^2+2\mk^2+m_{\pi^0}^2-\mkk^2-2m^2_{\rm cut}\right)\mkk c^2}{4M_{\Bs}q_\scr(\mkk)q_B(\mkk)},\label{eq:cthr}
\end{equation}
which reaches the physical region, $|\ckk|\leq 1$ when \mkk exceeds $\sim$1450 \mevcc. The error function entered in the definition of the acceptance accounts for the experimental resolution on the upper-limit value, $c_0(\mkk)$, through the  resolution parameter $\sigma_c$.
\begin{figure}[tb]
  \centering
  \begin{center}
    \includegraphics[width=18.5pc]{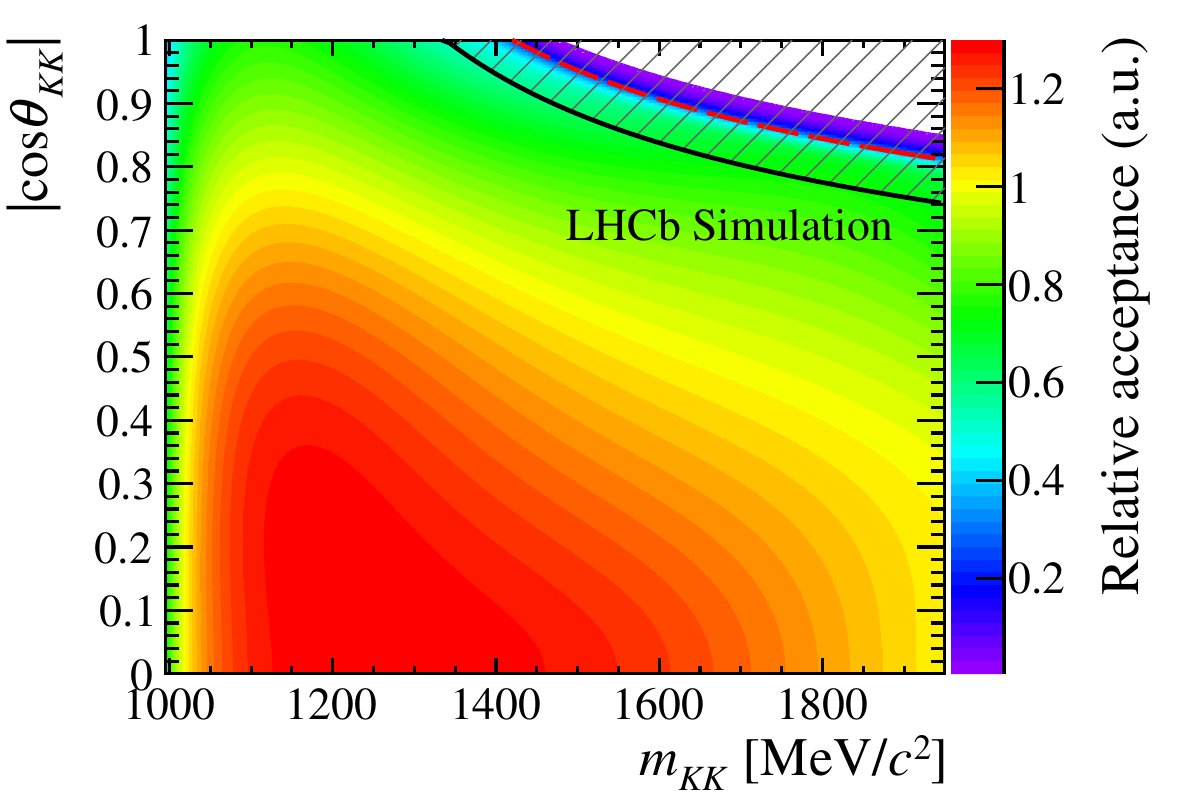}
    \includegraphics[width=18.5pc]{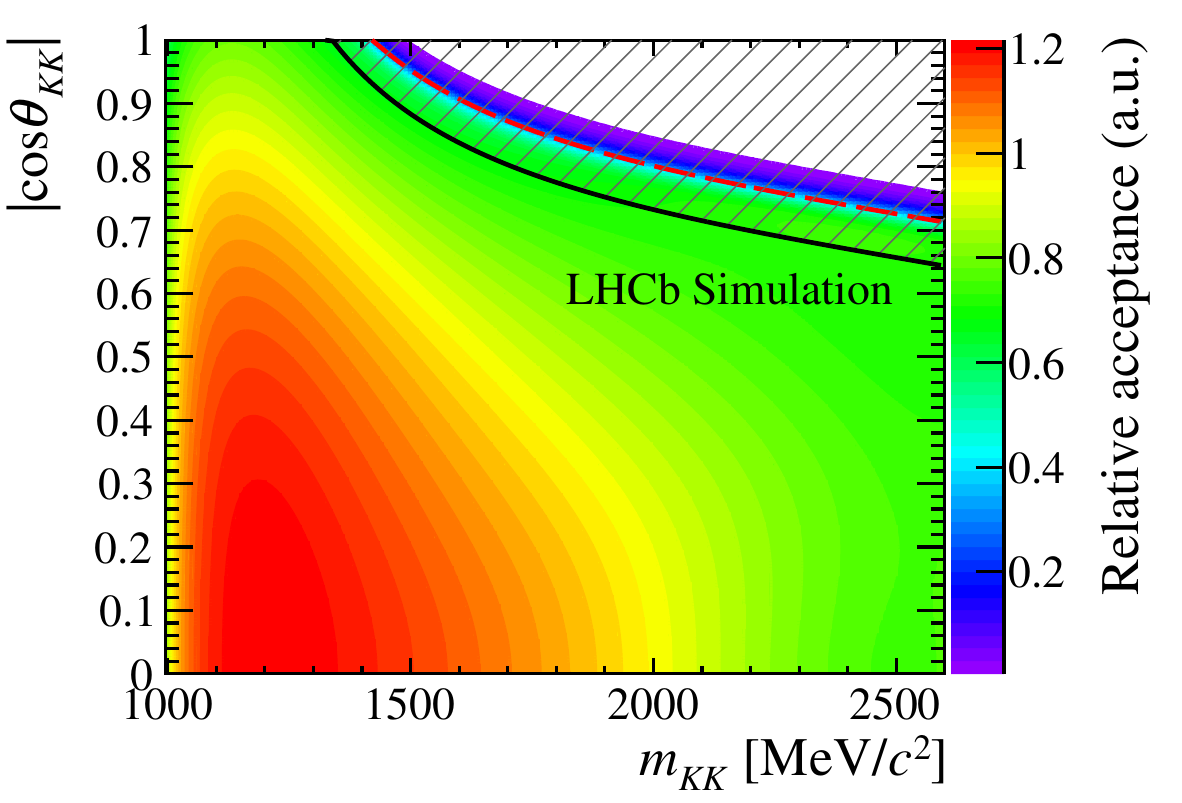}    
  \end{center}       
  \caption{Selection acceptance in the ($m_{KK},|\cos\theta_{KK}|$) plane for (left) Run 1 and (right) \mbox{Run 2}.
    The absolute normalisation $\varepsilon=1$ is arbitrarily set at the $\phi(1020)$ pole.
    The dashed-red curves indicate the kinematic boundary corresponding to the {anti-charm veto}, ${m_{K^\pm\gamma\to \pi^0}>2000}$~\mevcc. The hatched areas delimited by the solid-black curves indicate the fiducial acceptance cut applied to the data.}\label{fig:kkAcc2D}
\end{figure}

The acceptance is evaluated separately for the Run 1 and Run 2 data samples.
The set of acceptance parameters, $\vec\alpha=\{\alpha_{ij},\sigma_c\}$, as well as the mass pole, the width, the reconstructed mass resolution and the meson radius describing the mass shape of the resonances are let free to vary in the fit. The two-dimensional acceptance extracted from the simultaneous fit to the simulated samples is displayed in Fig.~\ref{fig:kkAcc2D} for Run 1 and Run 2.
The  helicity boundary resolution is found to be   $\sigma^{\fit}_c=(\val{2.17}{0.07})\times 10^{-2}$ for Run 1 and $\sigma^{\fit}_c=(\val{2.55}{0.03})\times 10^{-2}$ for  Run 2.
To avoid any systematic effect due to the modelling of the acceptance boundary on data, a fiducial cut, located 3$\sigma^{\fit}_c$ below the theoretical threshold, $|\ckk| < c_0(\mkk)-3\sigma_c^{\fit}$, is applied on the data sample, rejecting less than 1\% of the selected signal. The corresponding excluded acceptance is indicated by the hatched areas in Fig.~\ref{fig:kkAcc2D}. 

The fitted meson radius parameter, common to all the resonances in the fit, is found to be in very good agreement with the value used in the simulation. The  dikaon mass resolutions analytically included  in the relativistic Breit--Wigner model are found to be consistent with the nominal resolutions derived from the direct study of the $\Bs\to \phi\gamma$  and $\Bs\to f'_2(1525)\gamma$  simulated samples.
Small biases on the mass and width parameters due to reconstruction and selection effects are  measured in the fit to determine the acceptance function,
\begin{eqnarray}
\delta\mu^\simul_\fdv= \mu^\fit_\fdv-\mu^\simul_\fdv &=& (0.041\pm0.007~\eval{+0.004}{-0.014})\mevcc,\nonumber\\
\delta\Gamma^\simul_\fdv/\Gamma^\simul_\fdv &=& (2.1\pm 0.5\pm0.1)\perc,\nonumber\\
\delta\mu^\simul_\fqv                     &=& (0.3 \eval{0.1}\eval{0.1})\mevcc,\nonumber\\
\delta\Gamma^\simul_\fqv/\Gamma^\simul_\fqv  &=& (1.4\pm 0.3~\eval{+0.2}{-0.1})\perc,\nonumber
\end{eqnarray}
where the first uncertainties are due to the limited statistics of the simulated samples, and the second are systematic uncertainties obtained by varying the acceptance model and the simulation weighting procedures as discussed in Sect.~\ref{subsec:sWeightsyste}. These reconstruction biases derived from simulation are used to correct the  mass-shape parameters measured in data for the $\phi(1020)$ and the $f'_2(1525)$ resonances. 

\subsection{Background model}\label{ssec:background}
After \sPlot subtraction of the combinatorial background and the partially reconstructed $B$ decays, the background contamination is dominated by the misidentified  $\BdKpiGam$ and $\LbpKGam$ decays which are both  expected to contribute  at the level of a few percent in the signal region. A small and well localised contribution from  $B^0\to \Dzb(\KK)\pi^0$ decays with a high-energy neutral pion reconstructed as a photon is also expected. Other peaking contaminations, \eg charmless $\KK\pi^0$, are assumed to be small and therefore neglected in the nominal model. This assumption is addressed in the studies of systematic uncertainties. 

\begin{figure}[htb]
\centering
  \includegraphics[width=18.5pc]{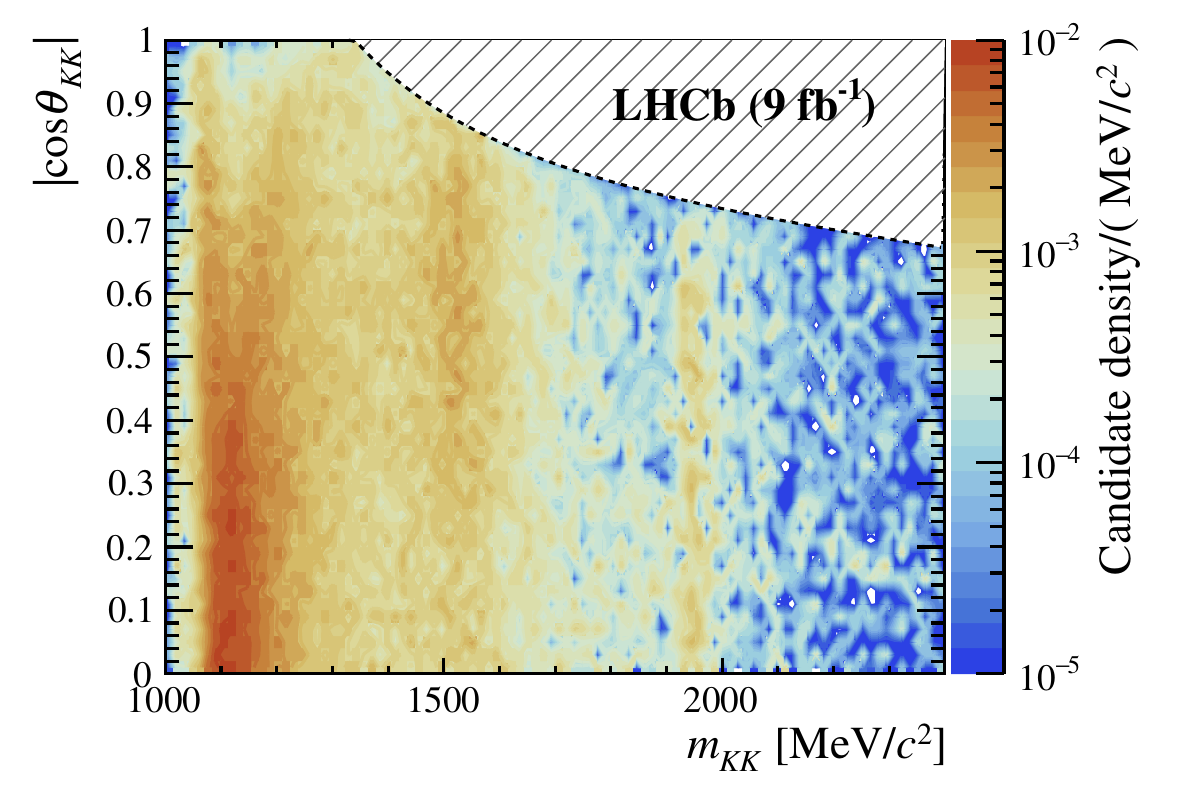}
  \includegraphics[width=18.5pc]{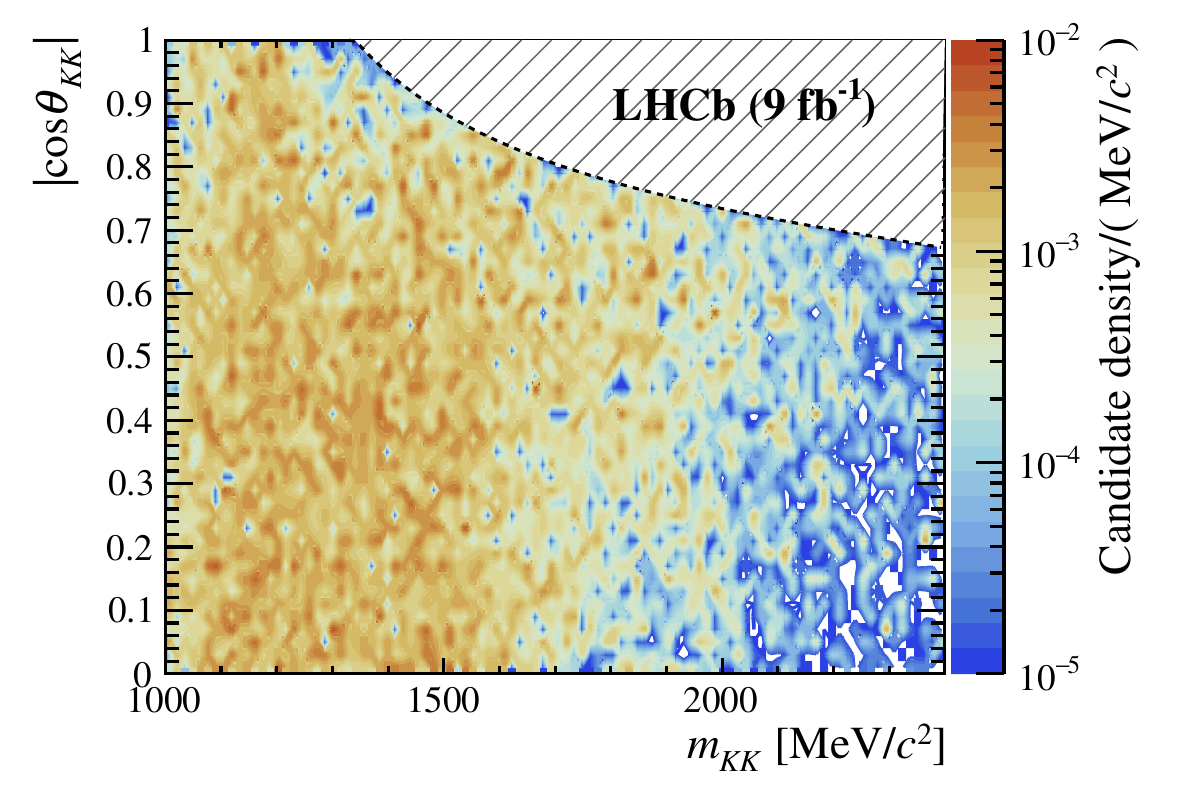}
  \caption{Reconstructed (left) $\Bd\to K^+\pi^-\gamma$ and (right) $\Lb\to pK^-\gamma$ candidates, projected on the ($m_{KK}$,$|\cos\theta_{KK}|$) plane. The dihadron mass and helicity observables are computed assigning the kaon mass to the pion and proton tracks. The mass-shifted contributions of $K^{*}(892)^{0}$, $K^{*}_2(1430)^{0}$ and $\Dzb\to K^+\pi^-$  are clearly visible in the projection of the $\Bd\to K^+\pi^-\gamma$ candidates.}\label{fig:ModelBkg} 
\end{figure}

The two-dimensional distribution of the $K^+\pi^-\gamma$ contamination in the (\mkk,\ackk) observables plane is modelled using a dedicated selection of a reconstructed and identified $K^+\pi^-\gamma$ data sample. The event reconstruction and  selection strictly reproduce the requirements discussed in Sect.~\ref{sec:Selection} for the \BsKKGam signal with an adapted exclusive criteria for the $K^+\pi^-$ dihadron identification. The \sPlot technique is used to extract the  $\BdKpiGam$ contribution, and the dihadron mass and helicity angle are both re-evaluated under the dikaon hypothesis, \ie by assigning a kaon mass to the pion candidate. The same procedure is applied to the baryonic $\LbpKGam$ decay, assigning the kaon mass to the proton candidate.
In addition to the mass substitution, $m_{\pi^+(p)}\to \mk$, correction weights derived from simulation are applied to the selected $K^+\pi^-\gamma$ and $pK^-\gamma$ data candidates to ensure their distributions correctly reproduce the corresponding misidentified contamination passing the exclusive dikaon identification requirements.
The resulting two-dimensional projections  of the  $24\times 10^4$ ($3\times 10^4$) $K^+\pi^-\gamma$  ($pK^-\gamma$) candidates, displayed in Fig.~\ref{fig:ModelBkg}, are used to build the background binned \pdfs.

The background-subtracted $\KKG$ sample possibly includes contaminations from high-energy neutral pions misidentified as photons.
The peaking $\left(K\pi^0\right)^\pm K^\mp$ decay modes are highly suppressed by the {anti-charm} veto.
The poorly known $\left(\KK\right)\pi^0$ charmless contamination is expected to be small and is treated as a source of systematic uncertainty. Other doubly misidentified contamination, for instance from $\left(K^+\pi^-\right)\pi^0$, are included in the data-derived description of the $\left(K^+\pi^-\right)\gamma$ contamination, discussed above.
The suppressed but well-localised  contamination from the charmed decay mode $\B^0\to\Dzb\pi^0$, with a visible branching fraction
$\BB(B^0\to \Dzb\pi^0)\cdot\BB(\Dzb\to \KK)=(\val{1.07}{0.06})\times 10^{-6}$, further suppressed by the neutral pion misidentification, is included in the amplitude model
as an incoherent scalar contribution with a uniform helicity distribution. The  $\Dzb\to \KK$ mass shape is described as a narrow Gaussian peak with a resolution of about 6.5 \mevcc, adjusted to the dikaon mass distribution around the $D^0$ mass selected in the $\KKG$ upper mass sideband.

\section{Amplitude Fit}\label{sec:AmpFit}
\subsection{Fit procedure and nominal isobar model}\label{ssec:fitmodel}

An unbinned extended likelihood fit of the nominal model is applied to the \sWeighted data sample, where the \sWeights are determined from a mass-fit to the combined Run~1 and Run~2 data. The model is adjusted  to data by minimizing the weighted negative log-likelihood function
defined as
\begin{equation}
-\mathrm{ln}\LL_\omega(\vec \rho)=-\sum_i^N {\omega}_i~ \mathrm{ln}\PP_i(\mkk,|\ckk|;{\vec \rho}\,) ,
\end{equation}
where $\PP_i(\mkk,|\ckk|;{\vec \rho}\,)$ is the model \pdf evaluated for the $i^{th}$ event observables given the set of parameters $\vec{\rho}$, and ${\omega}_i$ is the event weight derived from the  s${\cal P}$lot formalism. The \pdf defined in Eq.~\ref{eq:pdf} is applied to the whole data sample, combining Run 1 and Run 2 candidates, using the weighted average acceptance
\begin{equation}
  \varepsilon(\mkk,|\ckk|)=f_1\times \varepsilon_\textrm{Run1}(\mkk,|\ckk|)+(1-f_1)\times\varepsilon_\textrm{Run2}(\mkk,|\ckk|),
\end{equation}
where $f_1$ is the relative Run~1 signal yield normalized to the integrated acceptance ratio. 

As the signal \pdf normalisation is related to the  yield parameter $\NN_\textsc{s}$, one reference isobar contribution and it phase can be fixed by setting to unity the complex coefficient associated to the $\phi(1020)$ meson, $c_{\phi(1020)}=(1,0)$.
In the assumed limit of no $\Bs/\Bsb$ asymmetry, a reference phase can also be fixed for the incoherent even-spin subsystem. This is done by setting the phase of the  $f'_2(1525)$ amplitude factor to zero. With this convention, the even-spin complex phase is measured relative to the $f'_2(1525)$ state, and the phase of the odd-spin components is relative to the $\phi(1020)$ resonance.

\begin{table}[tb]
\centering
\caption{Selected states contributing to the  baseline isobar model. The nominal mass and width parameters used to model the resonant lineshapes are reported in the 3rd and 4th column\cite{PDG2022}(\cite{PhysRevD.80.031101}\note). When not specifically measured, the  branching fraction to the $\KK$ final state (5th column) is taken as half of the $\BB_{K\bar K}$ decay rate, neglecting any phase space effect for the heavy states far from the kinematic threshold. The last three columns indicate the measured isobar coefficients, $|c_\scr|$, the corresponding squared significance, $\chi^2_{|c_\scr|}$, and the increase in negative log-likelihood when the state is removed from the baseline model, \DNLL.}\label{tab:AmpModel} 
\resizebox{1.0\columnwidth}{!}{\small
\begin{tabular}{cc|ccc|ccc}
    \toprule 
    \small  State &\updown $J^{PC}$& $\mu_\scr$ \ummev   &   $\Gamma_\scr$ \ummev    &  $\BB_{\KK}$  \uperc      & $|c_\scr|$ \uten&\updown$\chi^2_{|c_\scr|}$& \DNLL  \\
    \midrule \fdv & \jp{1}{--}  & \val{1019.461}{0.016}   & \val{4.249}{0.013}         & \val{49.2}{0.5}             & 10 (fix)                    &  -   &  -         \\
    \fqv   & \jp{2}{++}  & \val{1517.4}{2.5}     & \val{86}{5}              & \val{43.8}{1.1}             & \val{4.16}{0.09}& 2270 &  -        \\
    \fsq   &\updown \jp{1}{--}& \val{1689}{12}\note   & \val{211}{24}\note            & seen                        & \val{2.40}{0.15}& 266 &  +304  \\
    \fds   & \jp{2}{++}  & \val{1275.5}{0.8}     & \val{186.6}{+2.2}{-2.5}  & \val{2.30}{+0.25}{-0.20}    & \val{1.07}{0.17}& 41  &  +18   \\
    \fdc   & \jp{3}{--}  & \val{1854}{7}         & \val{87}{+28}{-23}       & seen                        & \val{0.61}{0.16}& 14  &  +15   \\
    \fvd   & \jp{2}{++}  & \val{2011}{+62}{-76}  & \val{202}{+67}{-62}      & seen                        & \val{0.74}{0.18}& 16  &  +13   \\
    \fnr   & \jp{1}{--}  &  \multicolumn{3}{c|}{ - }                                                     & \val{0.79}{0.26}& 10 &  +17   \\
    \bottomrule
\end{tabular}}
\end{table}

The isobar model for the signal amplitude is built by selecting the possible contributions among the well-established unflavored isoscalar mesons that have been observed in the dikaon final state\cite{PDG2022}. Each  candidate  is accepted in the nominal model if it significantly improves the fit quality. Namely, the negative log-likelihood minimum is required to increase by more than $\DNLL=$12.5 units when the state is removed from the model, roughly indicating a $\sqrt{2\DNLL}=5$ standard deviations ($\sigma$) effect.\footnote{To better reflect their statistical interpretation, the quoted likelihood variations throughout the text implicitly include the global scaling factor
$\alpha = \frac{\sum_i^N {\omega}_i }{\sum_i^N {\omega}_i^2 }$ that aims at accounting for the  statistical dilution due to the \sPlot signal weights such that $\DNLL=\alpha\Delta\mathrm{ln}\LL_\omega$. The scale factor measured on the selected sample is $\alpha=0.67$.
}
In addition, the squared significance of the fitted isobar coefficient, $\chi^2_{|c_\scr|}=\left|{c_\scr}/{\sigma_{c_\scr}}\right|^2$, is required to exceed nine units to avoid selecting a poorly resolved amplitude that mostly improves the fit quality through its contribution to the interference pattern.
The selected states passing those criteria are summarized in Table~\ref{tab:AmpModel}, together with their statistical significances and the world average masses and widths used to parameterise their amplitude description. The free lineshape parameters for the $\fdv$ and $\fqv$ mesons are found to be consistent with the current measurements, and the $\fdv$ meson radius is measured as $\rphi=\val{1.01}{0.13}$\cgev. 

Besides the dominant contributions from the $\fdv$ and the $\fqv$ mesons, the fit indicates a high-significance contribution from the ($s\bar s$)-dominated vector meson, $\fsq$, with a fit fraction relative to the $\fdv$ of the order of $6\%$. The $\fds$ state, the isoscalar partner of the $\fqv$ tensor meson, is found to contribute at the level of $1\%$ to the overall amplitude, and the  J${}^\textrm{PC}=3^{--}$  $s\bar s$ candidate, $\fdc$, is measured with a relative fit fraction of the order of $0.3\%$. The amplitude fit also indicates a possible contribution from a heavy tensor state  around 2 \gevcc. Several  separate candidates are listed in Ref.~\cite{PDG2022} in that mass region, $f_2(1910)$, $f_2(1950)$, $f_2(2010)$ or $f_2(2150)$, which all contribute with a similar significance. The $\fvd$ state is slightly preferred by the fit, and is retained. 

The nominal model includes  a nonresonant component, \fnr, modelled as a pure P-wave uniformly distributed in mass  with a constant phase: $\AA\inr(\mkk,\tkk)=\dw{1}{10}(\tkk)$. The nonresonant amplitude is found to contribute at the level of $0.5\%$  with a statistical significance of $\DNLL=+17$. This contribution is, however, weakly resolved, and its significance is strongly correlated to the parameterisation of the other vector components. In particular, the significance of the nonresonant amplitude decreases either when the $\fsq$ state width parameter increases, or when the relativistic tail of the $\fdv$ meson increases at a low radius value. 

\begin{figure}[tb]
\centering
\includegraphics[width=25pc]{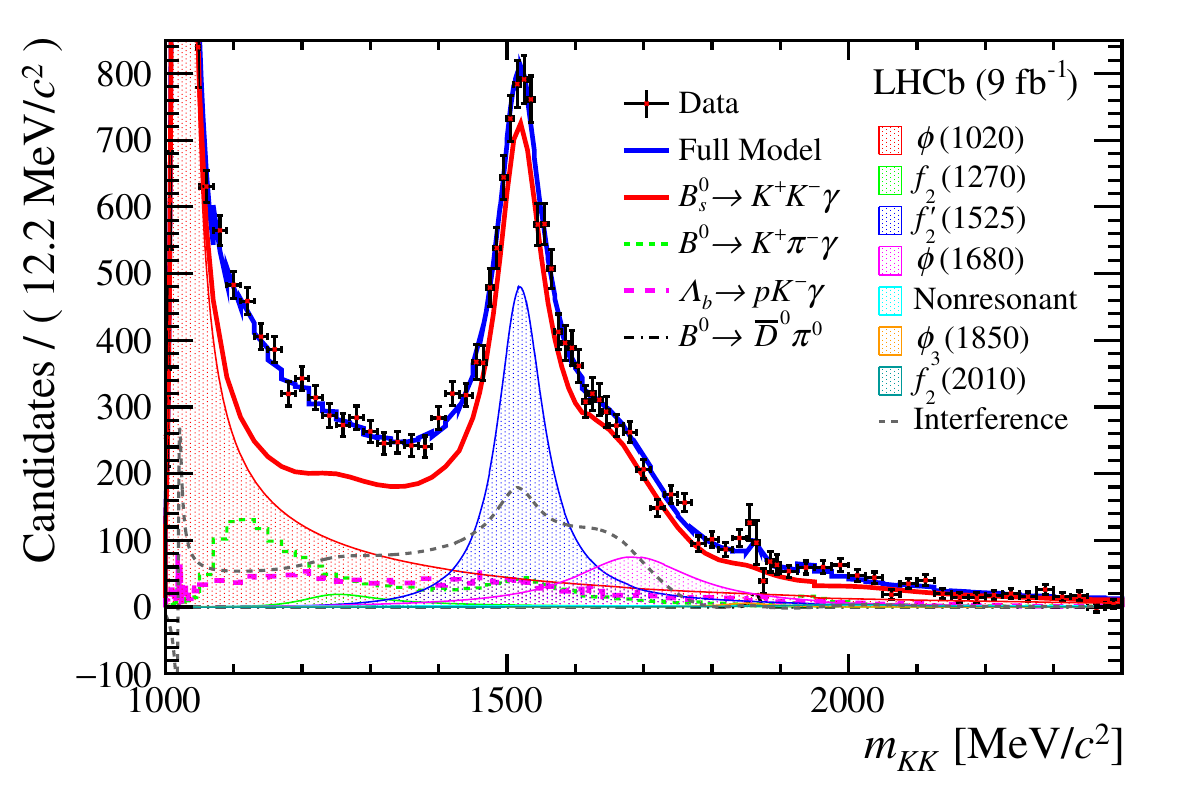}
\includegraphics[width=25pc]{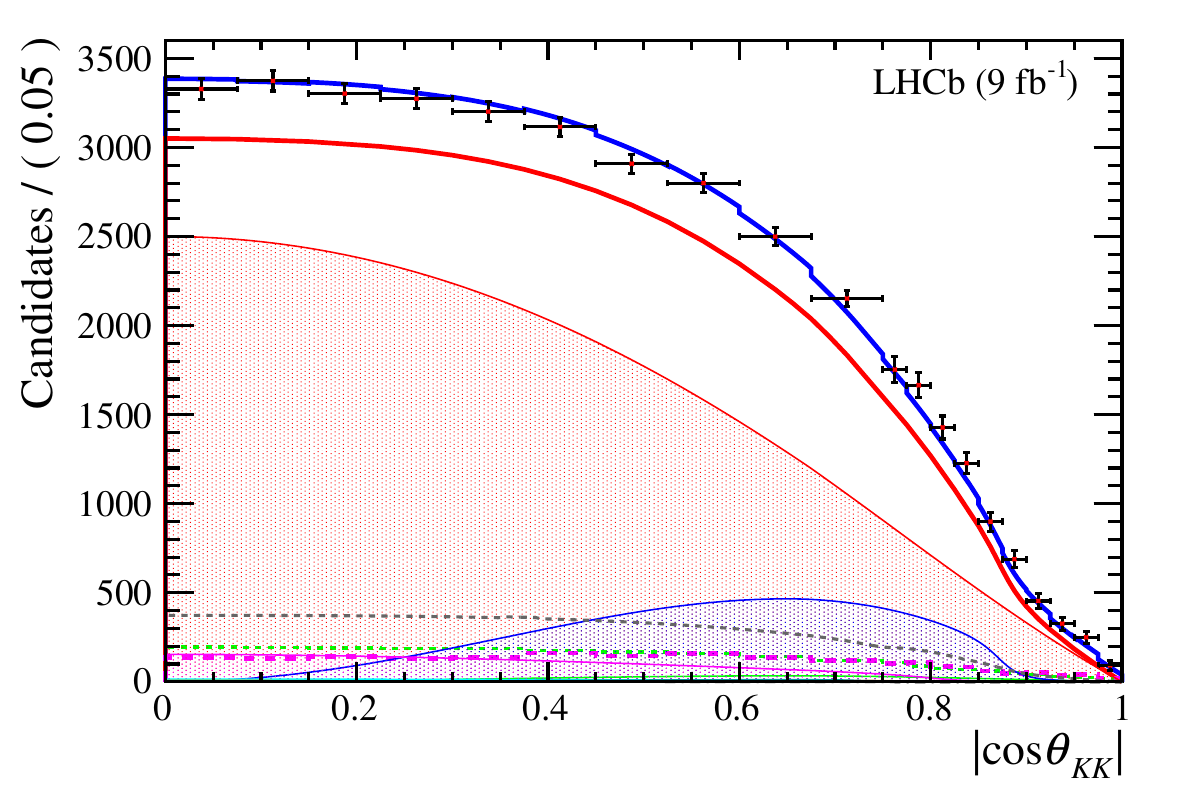}
  \caption{One-dimensional projection of the nominal fit on (top) $m_{KK}$ and (bottom) $|\cos\theta_{KK}|$ observables.
  Nonuniform binning is applied on both projections, with the bin width varying from 0.5\mevcc (in the $\phi(1020)$ region) to 20\mevcc for $m_{KK}$, and from 0.025 to 0.075 for $|\cos\theta_{KK}|$.  
  }\label{fig:FitA0}
\end{figure}

The nominal \pdf model depends on twenty free parameters: four overall normalisation yields, $\NN_\textsc{s}$, $\NN_\Bd$, $\NN_\Lb$ and $\NN_{D^0}$, parameterising the signal and backgrounds contributions to the data, six relative isobar factors $|c_\scr|$, their five relative phases $\delta_\scr$, the mass and width parameters of the dominant resonances, $\fdv$ and $\fqv$,  as well as the Blatt--Weisskopf radius parameter of the former. The Breit--Wigner parameters of other resonant states are fixed to the world average values reported in Table~\ref{tab:AmpModel}. The corresponding fit model projected on the mass and helicity observables, $\mkk$ and $|\ckk|$,  is shown in Fig.~\ref{fig:FitA0}. 
The fitted $\BsKKGam$ signal yield is found to be $\NN_\textsc{s}=(\val{44.4}{0.4})\times 10^3$. The fitted contamination of the misidentified backgrounds $\BdKpiGam$ and $\LbpKGam$ are \val{6.3}{0.7\%} and \val{5.4}{0.9\%}, respectively, in good agreement with the expectations derived from simulation studies discussed in Sect.~\ref{sec:Selection}. The residual background yield from the suppressed $B^0\to\Dzb(\KK)\pi^0$ decay is found to be $\NN_{D^0}< 110$, with 95\% confidence. \\

The individual contribution of each component $\scr$ to the $\BsKKGam$ amplitude model is measured by defining the fit fractions in the analysis mass range ${\mkk\in[2\mk,2400] \mevcc}$ as
\begin{equation}
  \FF_\scr =|c_\scr|^2\frac{\int_{0}^{+1}\int_{2\mk}^{2.4\mathrm{\gevcc}}|\AA_\scr(\mkk,|\ckk|)|^2 \diff\phi_3}{\int_{0}^{+1}\int_{2\mk}^{2.4\mathrm{\gevcc}}\sum_\textsc{p}\left|\sum_\scrp c_\scrp\cdot\AA_\scrp(\mkk,|\ckk|)\right|^2 \diff\phi_3} ,
\end{equation}
where {$\diff\phi_3=\JJ_3(\mkk)\diff\mkk\diff\!\ckk$} is the phase space volume. 
Although it is not required by the minimization process, each individual amplitude of the isobar model is normalised to unity,
\begin{equation}
  \int_{0}^{+1}\int_{2\mk}^{2.4\mathrm{\gevcc}}|\AA_\scrp(\mkk,|\ckk|)|^2 \diff\phi_3=1,
\end{equation}
to allow an easier interpretation of the associated isobar factor that directly provides the relative fit fractions normalized to $\fdv$: $\FF_\scr/\FF_\fdv=|c_\scr|^2$. 

\begin{figure}[htb]
\centering
  \includegraphics[width=21.pc]{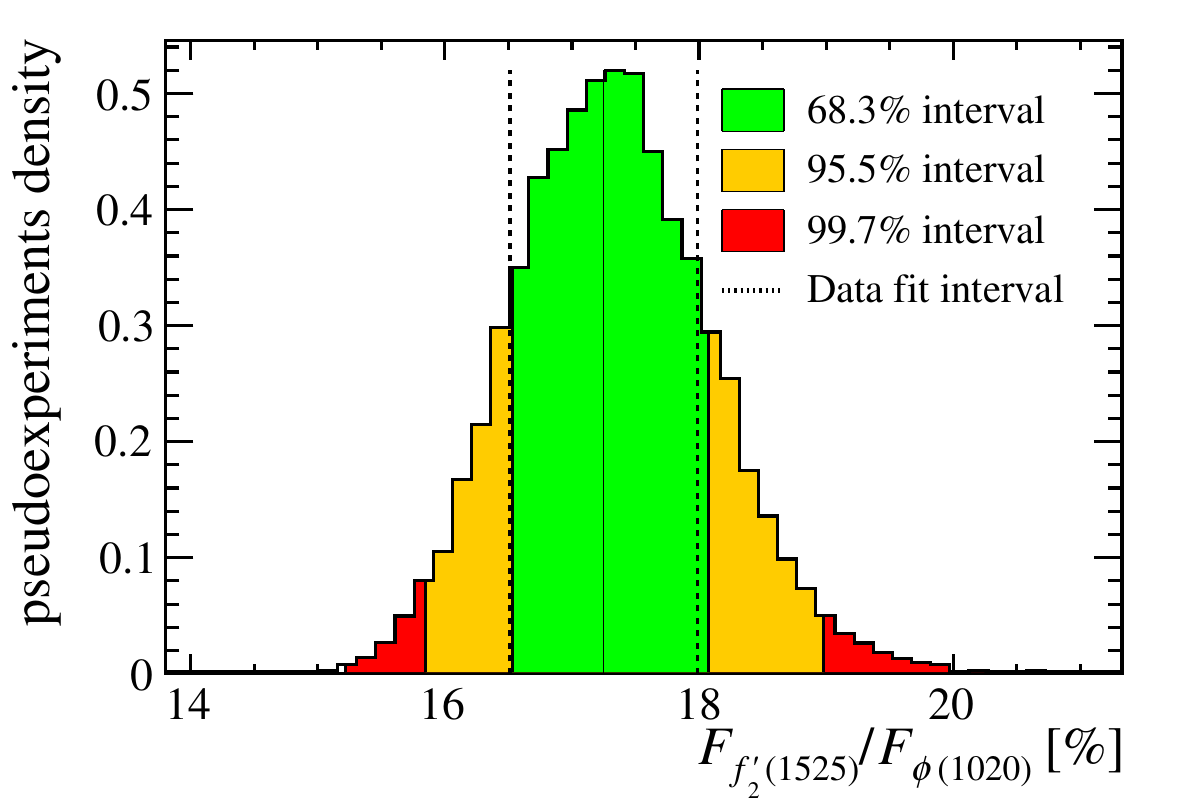}
  \caption{Distribution from pseudoexperiments of the $f'_2(1525)$ relative fit fraction normalized to the $\phi(1020)$ fraction.
    The one, two and three $\sigma$ asymmetric intervals, containing respectively 68.3$\%$, 95.5$\%$ and 99.7$\%$ of the pseudodata population on both sides of the distribution mean, are indicated by the green, orange and red coloured areas, respectively. The solid and dotted vertical lines indicate the central value and the statistical interval returned by the fit  to the data sample, evaluated using an asymptotically correct  approach.}\label{fig:FitToys1525}       
\end{figure}

To account for the statistical dilution due to the \sPlot weights, the asymptotically correct approach\cite{Langenbruch_2022,Dembinski:2021kim} is adopted to evaluate the statistical error on the fit parameters and on the derived fit fractions. The statistical coverage of the method is validated using a large sample of pseudoexperiments. For that purpose, a three-dimensional binned \pdf representation of the data sample observables (\mkkg, \mkk, $\ackk$) is prepared using an adaptative binning adjusted to the data density. Pseudodata samples are randomly generated from that \pdf with a random yield consistent with the size of the selected data sample. Each step of the analysis, including evaluating the s${\cal W}$eights, is applied to the generated pseudoexperiments.
As an illustration, the pseudodata dispersion of the $\fqv$ relative fit fraction, $\FF_{f'_2(1525)}/\FF_{\phi(1020)}$, compared with the statistical uncertainty derived from the fit using the asymptotically correct estimation of the parameters' covariance, is displayed in Fig.~\ref{fig:FitToys1525}.
The  statistical interval corresponding to 68.3\perc of the pseudodata population on both sides of the distribution maximum is found to be almost symmetrical and in  good agreement with the asymptotic error interval obtained from the fit to data. 

\subsection{Likelihood minima pattern}
Several distinct and almost degenerate \NLL minima separated by less than three units are found by exploring the isobar parameter space.
This quasi-degenerated behaviour can be explained by the poorly constrained interference pattern in the symmetrical $\BsKKGam$ decay where the odd- and  even-spin components form two incoherent amplitude systems. 

The even-spin subsystem consists of a largely dominant $\fqv$ amplitude surrounded by the two small $\fds$ and $\fvd$ contributions that may either interfere constructively or destructively with almost the same statistical significance.  A first set of four solutions separated by less than one log-likelihood unit, hereafter denoted $\LLA_i$ with $i\in\{0,1,2,3\}$, originate from the interference ambiguities in the even-spin subsystem, leaving the structure of the odd-spin system unchanged. The  overall minimum solution $\LLA_0$ ($\DNLL=0$) illustrated in Fig.~\ref{fig:FitA0} corresponds to the smallest fit fractions for each resonant state, along with positive interference. Other solutions with a larger amplitude and destructive interference, either for the $\fvd$  ($\LLA_1$, $\DNLL=0.1$) or the $\fds$ ($\LLA_2$,~$\DNLL=0.5$) or both ($\LLA_3$, $\DNLL=0.6$), are barely disfavoured. The overall contribution of the three spin-2 states, including their interference, is found to be $16.8\%$ identically for the four solutions $\LLA_i$ that only differ by their internal interference pattern.
The dominant contribution of the $\fqv$ state varies from 12\% with constructive interference for the preferred solution $\LLA_0$, to 20\% with negative interference for the solution $\LLA_3$. 

A similar ambiguity appears in the odd-spin system, which consists of the two well-separated vector resonances, $\fdv$ and $\fsq$, connected by the small nonresonant \mbox{P-wave}, $\fnr$, that can either interfere constructively or destructively in the $\fsq$ region. As a consequence, another set of four minima, denoted by $\LLB_i$ with $i\in\{0,1,2,3\}$, is obtained. It approximately replicates the $\LLA_i$ solutions for the even-spin subsystem and exhibits an alternative interference pattern for the vector components with a large $\fsq$ fit fraction of about $18\%$  along with large destructive interference, to be compared to $4\%$ with constructive interference for the solutions set $\LLA_i$. The small spin-3 amplitude, $\fdc$, does not induce any additional minimum pattern in the odd sector as its interference vanishes in the integration over the helicity observable due to the orthogonality of the Wigner d-functions. The four $\LLB_i$ solutions are modestly disfavored with a log-likelihood shift from $\DNLL=1.5$ to $2.9$ units.  The even-spin structure of each $\LLB_i$ solution is very close to that of the corresponding $\LLA_i$ one, with a consistent overall tensor fraction identical for the four solutions within 0.1\perc.
For illustration, the likelihood scans of the  $\fsq$ and the  $\fqv$ isobars in the vicinity of the solutions $\LLA_0$ and $\LLB_0$ are compared in Fig.~\ref{fig:FitScan}. The detailed amplitude structure of the eight quasi-degenerated solutions is summarized in Appendix~\ref{app:pattern}.
\begin{figure}[tb]
  \centering
  \includegraphics[width=18pc]{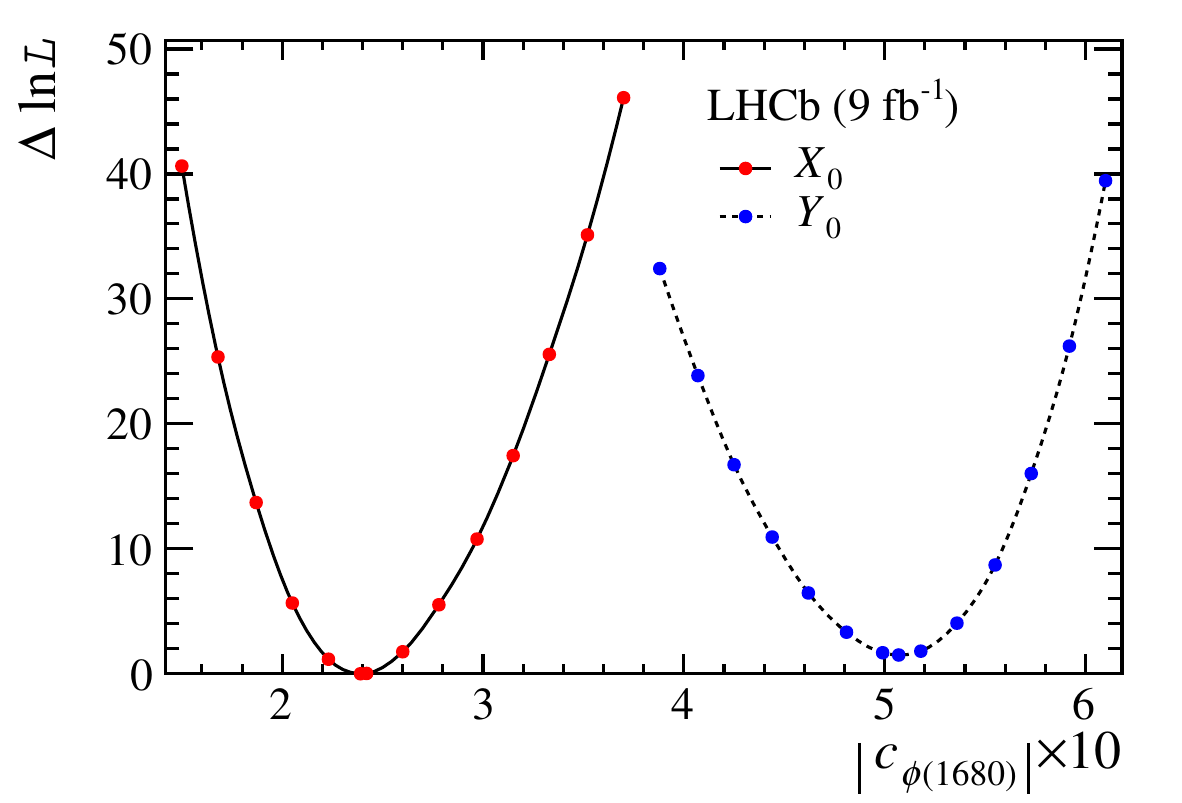}\includegraphics[width=18pc]{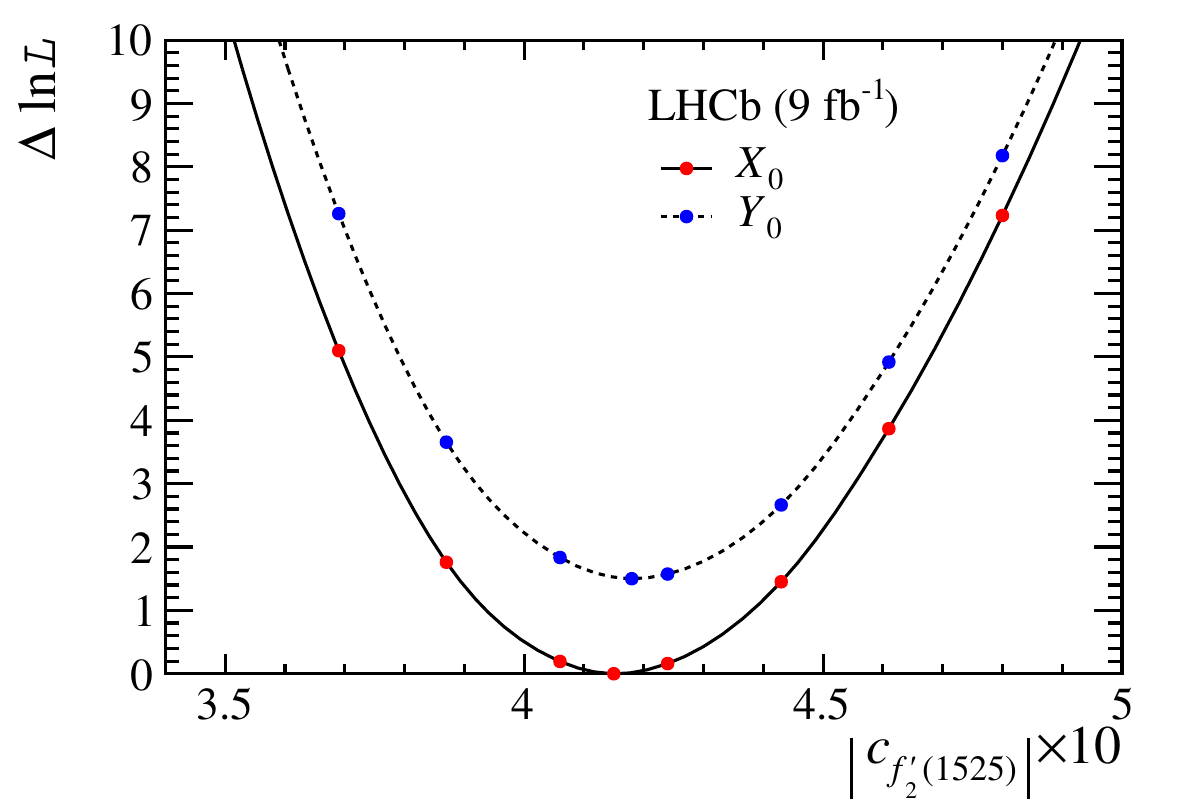}
  \caption{Negative log-likelihood scans as a function of (left) the  vector isobar $|c_{\phi(1680)}|$ and (right) the tensor isobar  $|c_{f'_2(1525)}|$  in the vicinity of the fit solutions $X_0$ and $Y_0$. The fit minima, well separated in the vector system, correspond to the same location in the tensor system.}\label{fig:FitScan}       
\end{figure}

The very large  $\fsq$ fit fraction of the $\LLB_i$ solutions is questionable, as this radial excitation of the ground state is expected to mainly decay into $\Kstarzb K$ and  $\phi\eta$ final states\cite{PhysRevD.86.012008,PDG2022}.
The small partial decay rates of the $\fsq$  to a kaon pair\cite{10.1007/BF01560440,PhysRevD.77.092002} disfavors a large contribution to the $(\KK)\gamma$ final state. The solutions $\LLB_i$ would imply an unlikely large $\Bs\to \fsq\gamma$ branching fraction ratio
\begin{equation}
  R_{\phi\gamma}=\frac{\BB(\Bs\to \fsq\gamma)}{\BB(\Bs\to \fdv\gamma)} \simeq 5.5\pm 0.9.\nonumber
\end{equation}
This presumption is also supported by the amplitude analysis of the  $\Bs\to \jpsi\KK$ decay\cite{LHCb-PAPER-2017-008} which reports the fit fraction $\FF^{\jpsi\KK}_{\fsq}=\val{4.0}{0.3}\estat\eval{0.3}\esyst \perc,$ very consistent with the value observed for the fit solutions $\LLA_i$. 

Similar arguments can be made for the orbitally excited tensor $\fqv$ and its almost decoupled isoscalar partner, $\fds$. The large $\fds$ fit fraction of the fit solutions $\LLA_2$ and $\LLA_3$ (or equivalently $\LLB_2$ and $\LLB_3$)  would imply the unlikely large ratio of branching fractions
\begin{equation}
  R_{f_2\gamma}=\frac{\BB(\Bs\to \fds\gamma)}{\BB(\Bs\to \fqv\gamma)} = 3.0\pm 0.3.\nonumber
\end{equation}
Although most of the quasi-degenerate fit solutions are disfavoured by external arguments, no definitive statement can  be derived from the amplitude fit to the current data,
given the numerical proximity of the negative log-likelihood minima. It is worth noting, however, that the tensor subsystem including the three states $f_2=\{\fds$, $\fqv$,$\fvd\}$ is found to contribute with an overall fit fraction of about $\val{17.0}\perc$ consistently for all the eight solutions. The $\fdv$ and $\fqv$ lineshape parameters as well as the signal
and background composition of the data sample are also found to be independent of the amplitude fit solutions.

\section{Systematic uncertainties}\label{sec:Syste}
The sources of systematic uncertainty can be organised into four main categories driven by the analysis steps: the uncertainties related to the $\KKG$ mass fit and the determination of the \sPlot weights, the uncertainties related to the parametrisation of the two-dimensional acceptance, the uncertainties related to the amplitude \pdf and the description of the backgrounds, and the uncertainties related to the nominal choices for the dikaon isobar model. 

\begin{table}[htpb]
\caption{Systematic uncertainties on the parameters of the amplitude fit: fit fractions, \FFr, relative fit fractions, \rFF, isobar phases, \dr, and  mass-shape parameters, $\mu_\textsc{r}$, $\Gamma_\textsc{r}$ and $\rphi$. The uncertainties due to the mass fit and the \sPlot weights, the two-dimensional acceptance definition, the amplitude fit \pdf and the isobar model are given from the 3rd to the 6th columns, respectively. The last column shows the overall uncertainties calculated as the quadratic sum of the individual sources.} \label{tab:Systematics}
\centering\resizebox{\columnwidth}{!}{%
\small
  \begin{tabular}{c|l|S[round-precision=1]S[round-precision=1]S[round-precision=1]S[round-precision=1]|S[round-precision=1]}
  \cmidrule(l){3-7}
   \multicolumn{2}{c}{}                   &  \mkkg fit & {Acceptance} & {Amp. fit} & {Isobar model} &  Total \\
\midrule
$\phi(1020)$              & \FFr \uperc    &\perval{+0.21}{-0.65} &\perval{+0.85}{-0.38} &\perval{+0.32}{-0.87} & \perval{+0.32}{-0.49}  & \perval{+0.99}{-1.19}\\
& \muR \umkev    &\peval{+5}{-6}        &\peval{+5}{-6}        &\peval{+5}{-4}        & \peval{+6}{-14}        & \peval{+11}{-16}\\
& \GamR \umkev   &\peval{+3}{-10}       &\peval{+8}{-7}        &\peval{+14}{-62}      & \peval{+20}{-82}       & \peval{+26}{-103}\\
& \rphi \ucgev   & \peval{+0.02}{-0.07} & \peval{+0.05}{-0.04} & \peval{+0.05}{-0.06} & \footnotesize\peval{0.09}           & \footnotesize\peval{0.12}   \\
\midrule
 ${ f_2(1270)}$           & \FFr \uperc    &\peval{+0.09}{-0.22} &\peval{+0.04}{-0.07} &\peval{+0.13}{-0.21} & \peval{+0.09}{-0.13}  & \perval{+0.19}{-0.34}\\
& \rFF \uperc    &\peval{+0.13}{-0.31} &\peval{+0.05}{-0.10} &\perval{+0.20}{-0.30} & \peval{+0.13}{-0.19}  & \perval{+0.28}{-0.49}\\
& \dr  \udeg     &\peval{+6.3}{-3.8}    &\peval{+1.1}{-0.9}     &\peval{+5.6}{-8.5}    & \peval{+23.1}{-14.1}   & \peval{+24.6}{-16.9}\\
 \midrule
  ${ f'_2(1525)}$         & \FFr \uperc    &\perval{+0.25}{-0.19} &\perval{+0.17}{-0.14} &\perval{+0.59}{-0.25} & \perval{+0.61}{-0.16}  & \perval{+0.90}{-0.38}\\
& \rFF \uperc    &\perval{+0.32}{-0.14} &\perval{+0.32}{-0.36} &\perval{+0.85}{-0.18} & \perval{+0.86}{-0.23}  & \perval{+1.29}{-0.49}\\
& \muR \ummev    &\peval{+0.3}{-0.8}    &\footnotesize\peval{0.2}           &\peval{+0.6}{-0.5}    & \peval{+1.2}{-1.7}     & \peval{+1.4}{-1.9}\\
& \GamR \ummev   &\peval{+0.6}{-0.9}    &\peval{+0.8}{-0.4}    &\peval{+2.0}{-0.6}    & \peval{+2.5}{-1.0}     & \peval{+3.4}{-1.5}\\
 \midrule
  ${ \phi(1680)}$         & \FFr \uperc    &\perval{+0.29}{-0.48} &\footnotesize\perval{0.18} &\footnotesize\perval{0.28} &\footnotesize \perval{0.47}  &\footnotesize \perval{0.70}\\ 
& \rFF \uperc    &\perval{+0.39}{-0.64} &\perval{+0.25}{-0.32} &\perval{+0.46}{-0.35} &\footnotesize \perval{0.70}  & \perval{+0.97}{-1.05}\\
& \dr  \udeg     &\peval{+2.7}{-3.1}    &\peval{+1.2}{-1.4}    &\peval{+3.9}{-1.7}    & \peval{+6.7}{-6.9}     & \peval{+8.3}{-7.9}\\
 \midrule
  ${ \phi_3(1850)}$       & \FFr \uperc    &\peval{+0.13}{-0.07} &\footnotesize\peval{0.03}         &\peval{+0.07}{-0.04}  & \peval{+0.12}{-0.10} & \peval{+0.19}{-0.13}\\
& \rFF \uperc    &\peval{+0.19}{-0.10} &\peval{+0.05}{-0.04} &\peval{+0.10}{-0.06}  & \peval{+0.17}{-0.14} & \perval{+0.28}{-0.19}\\
& \dr  \udeg     &\peval{+1.4}{-4.7}    &\peval{+2.7}{-1.5}    &\peval{+4.1}{-3.2}     & \peval{+11.5}{-10.5}    & \peval{+12.6}{-12.1}\\
 \midrule
  ${ f_2(2010)}$          & \FFr \uperc    &\peval{+0.08}{-0.01} &\peval{+0.02}{-0.07} &\peval{+0.14}{-0.07}  & \peval{+0.11}{-0.10} & \peval{+0.20}{-0.14}\\ 
& \rFF \uperc    &\peval{+0.11}{-0.01} &\peval{+0.02}{-0.10} &\peval{+0.21}{-0.10}  & \peval{+0.16}{-0.14} & \perval{+0.29}{-0.20}\\
& \dr  \udeg     &\peval{+17.1}{-15.9}   &\peval{+9.1}{-2.8}    &\peval{+20.5}{-11.7}  & \peval{+43.7}{-55.1}  & \peval{+52.1}{-58.6}\\  
 \midrule
  ${ (\textsc{kk})\inr}$  & \FFr \uperc    &\peval{+0.21}{-0.08}&\peval{+0.04}{-0.14}  &\peval{+0.23}{-0.12} & \peval{+0.12}{-0.09}  & \perval{+0.34}{-0.22}\\ 
& \rFF \uperc    &\peval{+0.31}{-0.11}&\peval{+0.07}{-0.21}  &\peval{+0.34}{-0.17} & \peval{+0.17}{-0.12}  & \perval{+0.50}{-0.32}\\
& \dr  \udeg     &\peval{+1.9}{-4.5}   &\peval{+1.7}{-2.8}     &\peval{+6.1}{-6.7}    & \peval{+5.5}{-3.9}     & \peval{+8.5}{-9.4}\\   
 \bottomrule
\end{tabular}
}
\end{table}

The systematic uncertainties are summarised in Table~\ref{tab:Systematics}, and details are provided in the following subsections.
The systematic uncertainties are presented as determined for the best-fit minimum, but their values are similar for other solutions.

\subsection{Mass fit and \sPlot weights.}\label{subsec:sWeightsyste}
The combinatorial $\KKG$ background and the partially reconstructed decays are statistically subtracted by applying \sPlot weights derived from a fit to the reconstructed invariant mass, $\mkkg$. The associated uncertainties reported in the third column of Table~\ref{tab:Systematics} are obtained by repeating the full amplitude analysis with alternative mass models to extract the \sPlot weights. The  fixed parameters that describe the tails on both sides of the signal peak have been varied and an alternative shape based on a first-order polynomial function has been tested for the combinatorial background.
Particular attention was paid to the modelling of the partially reconstructed backgrounds that extend into the signal mass region. The fixed parameters of the one-missing-pion  shape have been varied by several times their uncertainty as derived from simulation studies. Similar variations have been applied to the low-mass two-missing-pion  component that slightly overlaps with the signal peak. Additionally, the fit has been repeated in the reduced mass range, $\mkkg\in [5100,6400]\mevcc$, with this low-mass component removed from the fit model. To further check that the partially reconstructed background components are correctly subtracted in the signal mass region, the full data sample has been split into four bins of the dikaon mass, with very different levels of contamination: the $\phi(1020)$ region, \mkk$\in[1000,1100]$\mevcc, which is almost background free, the two intermediate regions  \mkk$\in[1100,1300]$\mevcc~and \mbox{\mkk$\in[1300,1525]$\mevcc}, where most of the partially reconstructed backgrounds accumulate, and the high dikaon mass region \mbox{\mkk$\in[1525,2400]$\mevcc}. A limited variation of the amplitude fit fractions is obtained when applying  \sWeights extracted in separate \mkk bins. The same procedure is applied to the helicity observable, splitting the data sample into four bins of $|\ckk|$ almost equally populated. The difference in the $\mkkg$ mass distribution over time has also been tested by splitting the data sample into data-taking periods. A small change in the amplitude fit results is obtained when extracting the \sWeights from separate mass fits per year of data taking or when splitting the data into two subsamples for Run 1 and \mbox{Run 2}. A simultaneous amplitude fit to the Run 1 and Run 2 data with separate acceptances and independent background contributions has alternatively been performed. The small difference with the nominal strategy is included in the budget of systematic uncertainties.

\subsection{Acceptance model }\label{subsec:AccSyste}
The  acceptance-related uncertainties reported in the fourth  column of Table~\ref{tab:Systematics} address the  acceptance model definition, the simulation corrections weights and the limited data samples used to derive the acceptance function. The robustness of the acceptance model has been tested against the choice of the considered simulated decays and using alternative parameterisations. The impact of the limited statistics has been evaluated by repeating the amplitude analysis with sets of acceptance parameters randomly generated according to their covariance. The uncertainties affecting the weights reproducing the dikaon and the photon identification efficiencies\cite{LHCb-PUB-2016-021,note-gammapi0calib} have been propagated to the acceptance estimate. Alternative weighting corrections\cite{Rogozhnikov_2016} have been tested for the kinematics distributions.

\subsection{Amplitude \pdf and residual backgrounds }\label{subsec:pdfSyste}
The signal \pdf defined in Eq.~\ref{eq:sigpdf} explicitly assumes an equal decay rate for \Bs and $\Bsb$ mesons. To check this assumption and the resulting interference cancellation, the amplitude fit is repeated in the unfolded (\mkk, $\ckk$) plane with an adapted \pdf for the signal component
\begin{equation}
  \PP_\textsc{s}(\mkk,\tkk)=\NN_\textsc{s}\left[ \frac{1-{a}}{2}\times\PP_{\Bs}+\frac{1+{a}}{2}\times\PP_\Bsb \right],
\end{equation}
allowing the asymmetry parameter, ${a}$, to vary freely. The measured helicity observable, \tkk, arbitrarily defined with respect to the direction of the positively charged kaon, leads to the flavour-dependent \pdf definitions
\begin{eqnarray}
  \PP_\Bs(\mkk,\tkk)&=&\varepsilon(\mkk,\tkk)\cdot\JJ_3(\mkk)\cdot|\sum_\scr c_\scr \cdot \AA_\scr(\mkk,\tkk)|^2 ,\\
  \PP_\Bsb(\mkk,\tkk)&=&\PP_\Bs(\mkk,\pi-\tkk),
\end{eqnarray}
where the coherent sum is over all contributions, regardless of their spin parity. 
In addition, the reference phase for the even-spin states, which is fixed to 0 for the $f'_2(1525)$ in the nominal fit, is allowed to vary to account for the non-exact cancellation of the interference between odd and even spin resonances.
The decay asymmetry and the $\fqv$ relative phase are found to be consistent with no asymmetry within two standard deviations
\begin{eqnarray}
  {a}&=&(-3.9\pm2.2)\times 10^{-2},\nonumber \\
  {\delta_\fqv}&=&\val{-46}{37}~\mathrm{deg.}\nonumber
\end{eqnarray}
The observed variations of the fit parameters that do not exceed 0.3\% for the fit fractions are added to the budget of the systematic uncertainties. 

A small impact on the amplitude fit is observed when varying the  \pdfs that describe the background contributions discussed in Sect.~\ref{ssec:background}.
The possible peaking contribution from charmless $B\to \KK\pi^0$ decays with a high-energy neutral pion misidentified as $\KKG$ is neglected in the nominal \pdf. The expected contamination, roughly estimated at the level of 0.5\% from simulations, is affected by large uncertainties as the actual resonant structure of this final state is poorly known. 
The final state most similar to the signal, $\Bs\to\phi\pi^0$,  has a predicted branching fraction of ${\cal O}(10^{-7})$, leading to an expected negligible contamination of ${\cal O}(0.1\perc)$.
The same final state in the $B^0$ decay is expected to be further suppressed, ${\cal O}(10^{-9})$, and constrained by the upper limit\cite{Belle:2012irf} provided by Belle of $\BR(B^0\to\phi\pi^0)<1.5\times 10^{-7}$. 
  Evidence at 3.5$\sigma$ for the $\Bd\to \KK\pi^0$ decay was reported by Belle\cite{Belle:2013azu},  with a relatively large branching fraction of $\BR=(\val{2.17}{0.60}\eval{0.24})\times 10^{-6}$. The  charged intermediate states, $K^{*\pm}K^\mp$, are removed from the analysis by the {anti-charm veto}. However, the unknown S-wave $(K^{+}K^-)\pi^0$ contribution might contaminate the $(\KK)\gamma$ signal at the level of ${\cal O}(1\perc)$. No clear statement about a possible intermediate $(K^{+}K^-)$ structure could be made with the Belle data, however, BaBar has reported\cite{PhysRevLett.99.221801} a possible observation of a broad structure peaking near $\mkk\sim 1500$ \mevcc~ in the corresponding charged decay, $B^+\to \KK\pi^+$. LHCb has performed the first amplitude analysis of this three-body charged mode\cite{LHCb-PAPER-2018-051}, and a good description of the data pattern in that mass region is also observed when including the contribution of a vector resonance compatible with  the $\rho(1450)$ hypothesis. 

The potentially neglected contamination from the charmless $\KK\pi^0$ decays, which may also include a residual contribution from $\KK\eta(\to\gamma\gamma)$ decays not fully accounted for in the partially reconstructed backgrounds, has been investigated by adding an incoherent S-wave contribution to the dikaon amplitude. Several possible models have been tested: a resonant S-wave contribution based on the scalar mesons $f_0(980)\pi^0$ or  $f_0(1500)\pi^0$, a resonant vector-scalar contribution based on $\phi(1020)\pi^0$ or $\rho^0(1450)\pi^0$, a resonant tensor-scalar contribution based on $f'_2(1525)\pi^0$, and a nonresonant scalar contributions $(\KK)\inr\pi^0$. The corresponding \pdfs are built assuming the same acceptance as for the $\KKG$ final state. A relativistic Breit--Wigner lineshape is used to describe the $f_0(1500)$ and the  $\rho(1450)$ resonances with their world average mass and width\cite{PDG2022}. A Flatté lineshape is used for the $f_0(980)$ resonance with the BES parameterisation\cite{Ablikim_2005}.  The angular dependencies are described by the relevant Wigner d-functions, $\dw{1}{00}$ ($\dw{2}{00}$) for the vector (tensor) resonances and $\dw{0}{00}=1$ for the scalar hypothesis.  Among the different investigated models, the largest contributions are obtained with the  $f_0(980)\pi^0$ hypothesis, ${\cal N}_{f_0\pi^0}/{\cal N}_\textsc{s}$=\val{5.1}{1.0~\perc}, and with the $\phi(1020)\pi^0$ hypothesis, ${\cal N}_{\phi\pi^0}/{\cal N}_\textsc{s}$=\val{1.2}{0.2~\perc}, improving the fit quality in both cases. The other considered scalar, vector or tensor contributions around 1.5 \gevcc, $f_0(1500)$, $\rho(1450)$ or $f'_2(1525)$, as well as the nonresonant S-wave hypothesis, are all found  to vanish in the fit. The possibly significant $\KK\pi^0$ contamination at low $\mkk$ mass, much larger than expected in particular for the very suppressed $\phi\pi^0$ final state hypothesis, may indicate an opportunistic improvement of the fit when adding these additional degrees of freedom. Systematic uncertainties are conservatively derived from the largest observed positive and negative variations of each isobar parameter when adding any of the above $\KK\pi^0$ hypotheses to the fit model.

\subsection{Isobar model}
The lineshape parameters are fixed to their world average measurements for the $f_2(1270)$, $\phi(1680)$, $\phi_3(1850)$ and $f_2(2010)$ components of the nominal isobar model, while they are  free to vary for the two dominant $\phi(1020)$ and $f'_2(1525)$ resonances. The amplitude fit has been repeated by varying each fixed parameter within its uncertainty range by $\pm 1\sigma$. The quadratic sum of the parameter variations, limited to few $0.1\perc$ at most for the fit fractions, are added to the systematic uncertainty. When allowed to vary freely, the lineshape parameters for the subdominant $\phi(1680)$ component are found to be
\begin{eqnarray}
  \mu_{\phi(1680)}    &=& \val{1688}{10} \estat  ~\mevcc,\nonumber\\
  \Gamma_{\phi(1680)} &=& ~~\val{264}{28}  \estat ~\mevcc,\nonumber
\end{eqnarray}
in  good agreement with the nominal value taken from the Belle measurement\cite{PhysRevD.80.031101} and consistent with the current world average estimate\cite{PDG2022}. Moreover, the statistical significance of the isobar coefficient for the nonresonant $\fnr$ and for the $\phi_3(1850)$ components are both reduced below the $3\sigma$ level with fit fractions measured as $\FF_\fnr=(0.34\pm0.27)\perc$ and $\FF_\fdc=(0.15\pm0.11)\perc$, respectively. 

A negligible impact on the fit fractions is observed when varying the nominal $\phi({1020})$ mass resolution, analytically included in the relativistic Breit--Wigner lineshape.
Similarly, for higher mass resonances, the  resolution has been varied by $\pm 50\%$ of the nominal value, inducing marginal variation on the amplitude fit parameters other than the free $\fqv$ width that varies accordingly.

The angular momentum in the radiative $B$ decay to a $J_\scr$-spin state can take  ${L_B=\{J_\scr-1,J_\scr,J_\scr+1\}}$ values. The amplitude fit is performed using the lowest value $J_\scr-1$ for all resonances. Repeating the fit by fixing the $L_B$ parameter to the other allowed eigenvalues leads to a negligible effect on the fit fractions. 

The meson radius that defines the centrifugal correction in the Blatt--Weisskopf form-factors is nominally set in the range $[0.5,3]$\cgev for the light resonances\cite{PhysRevD.5.624}. 
The $\fdv$ radius parameter $\rphi$  is left free to vary in the nominal fit, leading to a measured value of $r_\phi=\val{1.01}{0.13}$\cgev. 
The meson radius for heavier resonances is fixed to the nominal value $r_{f_2}=3.0$\cgev. The likelihood scan of this parameter indicates that large radii are preferred with no clear minimum of the negative log-likelihood up to very large unphysical values.  Associated systematics are derived from the maximal negative or positive  variations of each of the fit parameters in the wide radius range $r_{f_2}\in[2.0,10]$\cgev, corresponding to changes of the log-likelihood by $[+0.7,-3.0]$ units with respect to the nominal point.

\subsection{Other possible resonant states}\label{sec:OtherStates}
The nominal isobar model includes the isoscalar dikaon states that significantly improve the fit quality according to the selection criteria presented in Sect.~\ref{ssec:fitmodel}.
Several possible resonant candidates with poorly measured properties, and sometimes unclear spectroscopic classification, may additionally contribute in the high-mass region.
No large contribution is observed when adding any of the known candidates to the nominal model.
The most significant rejected candidate is the $f_4(2050)$ state, just below the significance threshold for inclusion. The measured partial rate, $\BR(f_4(2050)\to K\bar K)=\val{0.68}{+0.34}{-0.18}{\perc}$, indicates that this state would have a negligible contribution to the signal. This spin-4 meson is generally interpreted as an  almost decoupled $u\bar u+d\bar d$ isoscalar state. The yet unconfirmed associated  $s\bar s$  heavy partner, $f_4(2300)$, induces a rather significant fit quality improvement,  $\DNLL= -10$. The impact on the nominal fit fractions is small, and no additional systematic is included in the error budget. 

Assuming $SU(3)_F$ symmetry, the excited dikaon states are reasonably described by a  \mbox{$\rho$--$\omega$--$\phi$} model with an almost ideal singlet-octet mixing and can be interpreted as recurrences of the $\rho$--$\omega$--$\phi$ vector ground state. Except for the well measured $f_2(1270)$ isoscalar, the nominal amplitude model only includes $s\bar s$-dominant states.
The largest significance, albeit limited, when adding the partially decoupled isovector partner to the isobar model,  is observed for the wide and possibly mixed vector states $\rho(1450)$ and $\rho(1700)$ that both increase the fit quality by $\DNLL\approx -5$.
Including those states essentially has the same impact on the fit as the enlargement of the $\phi(1680)$ width discussed above.
A limited fit improvement, $\DNLL\approx -3$ units, is observed as well with the $a_2(1320)$ isovector partner of the orbitally excited $\fds$ and $\fqv$ states.
Including this resonance only affects the $f_2(1270)$ fit fractions due to its approximate mass and width degeneracy with its isoscalar partner. The measured $f_2(1270)$ fit fraction must be considered as possibly receiving contributions from both the quasi-degenerate isoscalar partners. No impact is observed with the $a_2(1700)$, possible partner of the $f_2(1640)$ and $f_2(1950)$ isoscalars, nor with the $\rho_3(1690)$ associated to the $\phi_3(1850)$ $s\bar s$ state.
As no clear evidence of contribution  is observed for any of the tested additional states, no additional systematic source is included in the error budget.

\section{Results and conclusions}\label{sec:Results}
An isobar amplitude analysis of the radiative \BsKKGam decay mode is performed in the  mass range $\mkk\in\left[2\mk,2400\right]\mevcc$.
The $\fdv$ vector meson, accounting for almost 70\perc of the amplitude, dominates the dikaon structure. 
Considering the resonant contributions of $\fds$, $\fqv$ and $\fvd$ states, the overall tensor contribution  to the amplitude is measured as
\begin{equation}
\FF_{\{f_2\}}=\val{16.8}{0.5}\stat\eval{0.7}\syst\perc\nonumber,
\end{equation}
mostly dominated by the $\fqv$ state. Several almost statistically equivalent solutions are obtained for the detailed resonant structure depending on whether the low contributing resonances interfere destructively or constructively with the dominant amplitudes.
The statistically preferred solution corresponds to the lowest values of all the individual fit fractions along with constructive interferences that contribute for 3.5\% and 8.1\% in the even-spin and odd-spin subsystems, respectively. The corresponding fit fractions are given in Table~\ref{tab:TabFitA0},  together with the measured relative phases. The first quoted uncertainties are  statistical and correspond to the 68.3\% intervals derived from pseudoexperiments and the second uncertainties are the associated systematic uncertainties.
The sum of partial fit fractions is less than unity due to the integrated interference. 
Larger individual fit fractions, up to $20\%$ for the $\fqv$ state, associated with large destructive interference, cannot be excluded.

\begin{table}[tb]
\centering
\caption{Absolute and relative fit fractions (in the  mass range $\mkk\in\left[2\mk,2400\right]\mevcc$) and the associated isobar phase for the best-fit solution. The first quoted uncertainties are  statistical and correspond to the 68.3\% intervals derived from pseudoexperiments,  while the second are systematic.} \label{tab:TabFitA0}
\small
\begin{tabular}{c|c|c|c}
\toprule
\small  State                    &  Fit fraction \uperc                       &  Relative fit fraction \uperc              &  Phase \udeg \\
\midrule
\fdv                     &  \val{70.3}{+0.9}{-1.0}\eval{+1.0}{-1.2}   &   100                                      &  ~~0 (fixed)     \\
\fds                     &  ~~~\val{0.8}{0.3}\eval{+0.2}{-0.3}           & ~~\val{1.2}{+0.4}{-0.3}\eval{+0.3}{-0.5}   &  ~\val{-55}{+13}{-17}\eval{+25}{-17}  \\
\fqv                     &  \val{12.1}{+0.6}{-0.5}\eval{+0.9}{-0.4}   &   \val{17.3}{+0.8}{-0.7}\eval{+1.3}{-0.5}  &  ~~0 (fixed)     \\
\fsq                     &  ~~\val{3.8}{+0.6}{-0.5}\eval{0.7}           &  ~~\val{5.4}{+0.9}{-0.6}\eval{+1.0}{-1.1}   &  \val{137}{+5}{-6}\eval{\,8} \\
\fdc                     &  ~~\val{0.3}{+0.2}{-0.1}\eval{+0.2}{-0.1}    &  ~~\val{0.4}{+0.3}{-0.2}\eval{+0.3}{-0.2}   &  ~\val{-61}{+16}{-13}\eval{+13}{-12} \\
\fvd                     &  ~~~\val{0.4}{0.2}\eval{+0.2}{-0.1}           & ~~\val{0.6}{+0.3}{-0.2}\eval{+0.3}{-0.2}   &  ~~~\val{43}{+30}{-24}\eval{+52}{-59}   \\
\fnr                     & ~~\val{0.5}{+0.4}{-0.2}\eval{+0.3}{-0.2}    & ~~\val{0.6}{+0.5}{-0.3}\eval{+0.5}{-0.3}   &  ~\val{165}{+6}{-16}\eval{\,9}   \\
\bottomrule
\end{tabular}
\end{table}

The branching fraction $\BR(\Bs\to \fqv\gamma)$ relative to $\BR(\Bs\to\fdv\gamma)$ can be derived from the fit fractions ratio as
 \begin{equation}
   \frac{\BR(\Bs\to\fqv\gamma)}{\BR(\Bs\to\fdv\gamma)}=\frac{\BR(\fdv\to \KK)}{\BR(\fqv\to \KK)}\cdot\frac{\FF_ {\fqv}}{\FF_{\fdv}}.
 \end{equation}
 Using the world average measurements reported in Table~\ref{tab:AmpModel} for the branching fraction of $\fdv$ and $\fqv$ into $K^+K^-$, the ratio
 \begin{equation}
   \frac{\BR(\Bs\to\fqv\gamma)}{\BR(\Bs\to\fdv\gamma)}= 0.194 \eval{+0.009}{-0.008}\estat \eval{+0.014}{-0.005} \esyst \eval{0.005}\ebr~\nonumber
 \end{equation} 
 is obtained for the statistically preferred fit solution that corresponds to the smallest value. The last uncertainty is associated with the ratio of measured branching fractions to the $\KK$ final state. 
  This result establishes the first observation of the radiative $\Bs$ decay to an orbitally excited  meson, $\Bs\to f'_2(1525)\gamma$, and the second radiative transition observed in the $\Bs$ sector\cite{Belle:2014sac}. 
  
  A relative branching ratio can similarly be derived  for the  $\fds$ tensor partner,
 \begin{eqnarray}
   \frac{\BR(\Bs\to \fds\gamma)}{\BR(\Bs\to\fdv\gamma)}= 0.25 \eval{+0.09}{-0.07}\estat \eval{+0.06}{-0.10} \esyst \eval{0.03}\ebr,~\nonumber
 \end{eqnarray}
 which possibly includes the contribution from its quasi-degenerate isovector partner, $a_2(1320)$.
 The relative branching fraction of the $\phi(1680)\to \KK$ contribution is measured as
 \begin{eqnarray}
   \frac{\BR(\Bs\to \fsq\gamma)}{\BR(\Bs\to\fdv\gamma)}\times\BR(\fsq\to \KK)= 0.026 \eval{+0.004}{-0.003}\estat\eval{0.005}\esyst.\nonumber
\end{eqnarray}

 
 The mass and width of the $f'_2(1525)$ meson are measured, identically for all the almost degenerate solutions, as
\begin{eqnarray}
  \mu_{f'_2(1525)}    &=& \val{1521.8}{1.7}  \estat~\eval{+1.4}{-1.9}\esyst ~\mevcc,\nonumber\\
  \Gamma_{f'_2(1525)} &=& ~~~\val{79.3}{3.5} \estat~\eval{+3.3}{-1.5}\esyst ~\mevcc,\nonumber 
\end{eqnarray}
in good agreement with the current world average\cite{PDG2022} and  with the previous LHCb measurement\cite{LHCb-PAPER-2017-008}.
The precise measurement of the $\phi(1020)$  parameters gives
\begin{eqnarray}
  \mu_{\phi(1020)}    &=& \val{1019.50}{0.02}\estat\eval{0.02}\esyst ~\mevcc,\nonumber\\
  \Gamma_{\phi(1020)} &=& ~~~~\val{4.36}{0.05}  \estat~\eval{+0.03}{-0.10} \esyst \mevcc,\nonumber 
\end{eqnarray}
consistent with their current world average within 1.5 standard deviations.
The corresponding Blatt--Weisskopf radius parameter is measured to be
\begin{equation}
  \rphi=\val{1.0}{0.2}\estat \eval{0.1}\esyst ~~\cgev.\nonumber
\end{equation}


\section*{Acknowledgements}
%
%
\noindent We express our gratitude to our colleagues in the CERN
accelerator departments for the excellent performance of the LHC. We
thank the technical and administrative staff at the LHCb
institutes.
We acknowledge support from CERN and from the national agencies:
CAPES, CNPq, FAPERJ and FINEP (Brazil); 
MOST and NSFC (China); 
CNRS/IN2P3 (France); 
BMBF, DFG and MPG (Germany); 
INFN (Italy); 
NWO (Netherlands); 
MNiSW and NCN (Poland); 
MCID/IFA (Romania); 
MICINN (Spain); 
SNSF and SER (Switzerland); 
NASU (Ukraine); 
STFC (United Kingdom); 
DOE NP and NSF (USA).
We acknowledge the computing resources that are provided by CERN, IN2P3
(France), KIT and DESY (Germany), INFN (Italy), SURF (Netherlands),
PIC (Spain), GridPP (United Kingdom), 
CSCS (Switzerland), IFIN-HH (Romania), CBPF (Brazil),
and Polish WLCG (Poland).
We are indebted to the communities behind the multiple open-source
software packages on which we depend.
Individual groups or members have received support from
ARC and ARDC (Australia);
Key Research Program of Frontier Sciences of CAS, CAS PIFI, CAS CCEPP, 
Fundamental Research Funds for the Central Universities, 
and Sci. \& Tech. Program of Guangzhou (China);
Minciencias (Colombia);
EPLANET, Marie Sk\l{}odowska-Curie Actions, ERC and NextGenerationEU (European Union);
A*MIDEX, ANR, IPhU and Labex P2IO, and R\'{e}gion Auvergne-Rh\^{o}ne-Alpes (France);
AvH Foundation (Germany);
ICSC (Italy); 
GVA, XuntaGal, GENCAT, Inditex, InTalent and Prog.~Atracci\'on Talento, CM (Spain);
SRC (Sweden);
the Leverhulme Trust, the Royal Society
 and UKRI (United Kingdom).

\section*{Appendices}
\appendix
\section{Fit minima pattern}\label{app:pattern}
\begin{table}[tb]
  \caption{ Fit fractions , $\FF_\scr$, and interference fractions, ${\cal I}({\scr_1},{\scr_2})$, in the  mass range ${\mkk\in\left[2\mk,2400\right]\mevcc}$, for (top) the even-spin components and (bottom) the odd-spin components for the fit minima (left) $\LLA_i$ and (right) $\LLB_i$.
    The total integrated interference, ${\cal I}_{\{\scrp\}}$, and the overall fit fraction, $\FF_{\{\scrp\}}$  for the odd- ($\scr_-=\phi$) and the even-spin ($\scr_+=f_2$) subsystems
    are emphasised.} \label{tab:MirrorsA}
\centering\resizebox{1.\columnwidth}{!}{%
  \centering\small
  \begin{tabular}{c|cccccccc|cccccccc}
    \toprule
    \updown$\mathbf\DNLL$
    &\multicolumn{2}{c}{\bf0.0 ($\LLA_0$)}&\multicolumn{2}{c}{\bf0.1 ($\LLA_1$)}&\multicolumn{2}{c}{\bf0.5 ($\LLA_2$)}&\multicolumn{2}{c|}{\bf0.6 ($\LLA_3$)} 
    &\multicolumn{2}{c}{\bf1.5 ($\LLB_0$)}&\multicolumn{2}{c}{\bf1.6 ($\LLB_1$)}&\multicolumn{2}{c}{\bf2.5 ($\LLB_2$)}&\multicolumn{2}{c}{\bf2.9 ($\LLB_3$)} \\
    \midrule
    \multicolumn{1}{c|}{\gray$\bf{J^{PC}=2^{++}}$}
    & \multicolumn{8}{c|}{\gray$\FF_{\{f_2\}}$=\valbf{16.8}{0.5}\perc}
    & \multicolumn{8}{c}{\gray$\FF_{\{f_2\}}$=\valbf{17.3}{0.6}\perc} \\
    \midrule
    $\FF_\fqv$&\multicolumn{2}{c}{~\val{12.1}{0.5}}&\multicolumn{2}{c}{~\val{13.8}{0.6}}&\multicolumn{2}{c}{~\val{17.9}{0.7}}&\multicolumn{2}{c|}{~\val{20.4}{0.8}}
                    &\multicolumn{2}{c}{~\val{12.2}{0.5}}&\multicolumn{2}{c}{~\val{13.9}{0.6}}&\multicolumn{2}{c}{~\val{18.3}{0.8}}&\multicolumn{2}{c}{~\val{20.8}{0.8}}\\  
    $\FF_\fds$&\multicolumn{2}{c}{~~\val{ 0.8}{0.3}}&\multicolumn{2}{c}{~~\val{ 0.9}{0.3}}&\multicolumn{2}{c}{~~\val{ 2.9}{0.5}}&\multicolumn{2}{c|}{~~\val{ 3.2}{0.6}}
                    &\multicolumn{2}{c}{~~\val{0.9}{0.3}}&\multicolumn{2}{c}{~~\val{ 0.9}{0.3}}&\multicolumn{2}{c}{~~\val{ 3.1}{0.5}}&\multicolumn{2}{c}{~~\val{ 3.4}{0.6}}\\
    $\FF_\fvd$&\multicolumn{2}{c}{~~\val{0.4}{0.2}}&\multicolumn{2}{c}{~~\val{ 3.7}{0.5}}&\multicolumn{2}{c}{~~\val{ 0.5}{0.2}}&\multicolumn{2}{c|}{~~\val{ 4.1}{0.8}}
                    &\multicolumn{2}{c}{~~\val{ 0.5}{0.2}}&\multicolumn{2}{c}{~~\val{ 4.3}{0.6}}&\multicolumn{2}{c}{~~\val{ 0.6}{0.2}}&\multicolumn{2}{c}{~~\val{ 4.8}{0.6}}\\
\updown${\cal I}({\fqv,\fds})$
                   &\multicolumn{2}{c}{\val{+2.4}{0.4}} &\multicolumn{2}{c}{ \val{+2.3}{0.6}}&\multicolumn{2}{c}{ \val{-5.7}{0.7}}&\multicolumn{2}{c|}{\val{-6.3}{0.6}}
                   &\multicolumn{2}{c}{\val{+2.6}{0.4}} &\multicolumn{2}{c}{ \val{+2.7}{0.5}}&\multicolumn{2}{c}{ \val{-6.1}{0.7}}&\multicolumn{2}{c}{\val{-6.5}{0.6}}\\
\updown${\cal I}({\fqv,\fvd})$
                   &\multicolumn{2}{c}{\val{+0.7}{0.4}} &\multicolumn{2}{c}{ \val{-3.1}{0.6}}&\multicolumn{2}{c}{ \val{+1.7}{0.4}}&\multicolumn{2}{c|}{\val{-4.9}{0.4}}
                   &\multicolumn{2}{c}{\val{+0.8}{0.4}} &\multicolumn{2}{c}{ \val{-3.5}{0.6}}&\multicolumn{2}{c}{ \val{+1.9}{0.3}}&\multicolumn{2}{c}{\val{-5.3}{0.4}}\\

\updown${\cal I}({\fvd,\fds})$
                   &\multicolumn{2}{c}{\val{+0.3}{0.1}} &\multicolumn{2}{c}{ \val{-0.7}{0.4}}&\multicolumn{2}{c}{ \val{-0.5}{0.3}}&\multicolumn{2}{c|}{\val{+0.3}{0.5}}
                   &\multicolumn{2}{c}{\val{+0.4}{0.1}} &\multicolumn{2}{c}{ \val{-1.0}{0.3}}&\multicolumn{2}{c}{ \val{-0.5}{0.3}}&\multicolumn{2}{c}{\val{+0.1}{0.4}}\\
\gray${\cal I}_{\{f_2\}}$
                   &\multicolumn{2}{c}{\gray\valb{+3.5}{0.5}}&\multicolumn{2}{c}{\gray\valb{-1.5}{0.5}}&\multicolumn{2}{c}{\gray\valb{-4.6}{0.9}}&\multicolumn{2}{c|}{\gray\valb{-11.0}{1.1}}
                   &\multicolumn{2}{c}{\gray\valb{+3.8}{0.5}}&\multicolumn{2}{c}{\gray\valb{-1.8}{0.5}}&\multicolumn{2}{c}{\gray\valb{-4.7}{0.8}}&\multicolumn{2}{c}{\gray\valb{-11.8}{1.0}}\\
\midrule
 \multicolumn{1}{c|}{\gray$\bf{J^{PC}=(1,3)^{--}}$}
   & \multicolumn{8}{c|}{\gray$\FF_{\{\phi\}}$=\valbf{83.2}{0.5}\perc} 
   & \multicolumn{8}{c}{\gray$\FF_{\{\phi\}}$=\valbf{82.7}{0.6}\perc} \\
\midrule
    $\FF_\fdv$
    &\multicolumn{8}{c|}{ \val{70.4}{1.0}}
    &\multicolumn{8}{c}{ \val{71.1}{0.7}}\\
    $\FF_\fsq$
    &\multicolumn{8}{c|}{ ~~\val{ 4.0}{0.5}}
    &\multicolumn{8}{c}{ \val{18.2}{1.0}}\\
    $\FF_\fdc$
    &\multicolumn{8}{c|}{ ~~\val{ 0.3}{0.1}}
    &\multicolumn{8}{c}{ ~~\val{ 0.2}{0.1}}\\
    $\FF_\fnr$
    &\multicolumn{8}{c|}{ ~~\val{ 0.4}{0.3}}
    &\multicolumn{8}{c}{ ~~\val{ 0.3}{0.2}}\\
\updown${\cal I}({\fdc,\fsq)}$
           &\multicolumn{8}{c|}{ \val{+3.9}{0.3}}
           &\multicolumn{8}{c}{ \val{-7.8}{0.5}}\\
\updown${\cal I}({\fdv,\fnr})$
           &\multicolumn{8}{c|}{ \val{+4.0}{1.2}}
           &\multicolumn{8}{c}{ \val{+2.7}{1.0}}\\
\updown${\cal I}({\fsq,\fnr})$
           &\multicolumn{8}{c|}{ \val{+0.3}{0.2}}
           &\multicolumn{8}{c}{ \val{-2.0}{0.8}}\\
\updown${\cal I}({\fdc,\phi_1})$
           &\multicolumn{8}{c|}{~~0}&\multicolumn{8}{c}{~~0}\\
\updown${\cal I}_{\{\phi\}}$
           &\multicolumn{8}{c|}{\gray\valb{+8.1}{1.0}}
           &\multicolumn{8}{c}{\gray\valb{-7.2}{0.6}}\\
\bottomrule
\end{tabular}
}
\end{table}

\indent
Table \ref{tab:MirrorsA} reports the fit fractions and the integrated interference fractions, ${{\cal I}({\scr_1},{\scr_2})=\int\left[|{\cal A}_{\scr_1}+{\cal A}_{\scr_2}|^2-|{\cal A}_{\scr_1}|^2-|{\cal A}_{\scr_2}|^2\right]}$, for the quasi-degenerated fit minima.
The corresponding isobar coefficients and relative phases are given in Table~\ref{tab:MirrorsB}.
The quoted symmetrical errors correspond to the asymptotically correct estimation of the statistical intervals.

\begin{table}[tb]
\caption{Isobar coefficients and phases for (top) the fit minima $\LLA_i$ and (bottom) $\LLB_i$.} \label{tab:MirrorsB}
\centering\resizebox{1.0\columnwidth}{!}{%
\small\centering
\begin{tabular}{ccc|cc|cc|cc}
\cmidrule(l){2-9}
&\multicolumn{2}{c|}{$\mathbf\LLA_0$}  &\multicolumn{2}{c|}{$\mathbf\LLA_1$ }&\multicolumn{2}{c|}{$\mathbf\LLA_2$}&\multicolumn{2}{c}{$\mathbf\LLA_3$}\\
&\updown$|c_\scr|$ \uten &\dr\udeg      &$|c_\scr|$ \uten &\dr\udeg           &$|c_\scr|$ \uten  & \dr\udeg         &$|c_\scr|$ \uten & \dr\udeg    \\
\midrule
\multicolumn{1}{c|}{\fdv}& \val{10}             & 0& \val{10}             & 0& \val{10}             & 0& \val{10}             & 0\\ 
\multicolumn{1}{c|}{\fsq}& \val{2.40}{0.15}     &\val{138}{4}  &\val{2.39}{0.14}  &\val{138}{4}       &\val{2.38}{0.14}  &\val{138}{4}      & \val{2.38}{0.14} &\val{137}{4}\\
\multicolumn{1}{c|}{\fdc}& \val{0.61}{0.16}     &~\val{-61}{13} &\val{0.61}{0.15}  &~\val{-61}{12}      &\val{0.60}{0.16}  &~\val{-61}{13}     & \val{0.60}{0.16} &~\val{-61}{13}\\
\multicolumn{1}{c|}{\fnr}& \val{0.79}{0.26}     &\val{165}{8}  &\val{0.80}{0.21}  &\val{165}{6}       &\val{0.81}{0.21}  &\val{165}{6}      & \val{0.81}{0.21} &\val{164}{7}\\
\multicolumn{1}{c|}{\fqv}& \val{4.16}{0.09}     &      0       &\val{4.43}{0.10}  &   0               &\val{5.04}{0.12}  &  0               & \val{5.38}{0.11} &   0\\
\multicolumn{1}{c|}{\fds}& \val{1.07}{0.17}     &\val{-55}{14} &\val{1.11}{0.17}  &\val{-72}{16}      &\val{2.04}{0.19}  &\val{146}{7}      & \val{2.13}{0.19} &\val{130}{6}\\
\multicolumn{1}{c|}{\fvd}& \val{0.74}{0.18}     &~~\val{42}{26}  &\val{2.29}{0.17}  &\val{-122}{10}     &\val{0.86}{0.20}  &~~\val{93}{28}      & \val{2.43}{0.18} &~~\val{68}{8}\\
\midrule
           &\multicolumn{2}{c|}{$\mathbf\LLB_0$}  &\multicolumn{2}{c|}{$\mathbf\LLB_1$ }&\multicolumn{2}{c|}{$\mathbf\LLB_2$}&\multicolumn{2}{c}{$\mathbf\LLB_3$}\\
\midrule
\multicolumn{1}{c|}{\fdv}& \val{10}             & 0& \val{10}             & 0& \val{10}             & 0& \val{10}             & 0\\ 
\multicolumn{1}{c|}{\fsq}& \val{5.06}{0.15}     &\val{-135}{3} &\val{5.07}{0.15}  &\val{-135}{3}       &\val{5.09}{0.15} &\val{-134}{3}     & \val{5.09}{0.15} &\val{-134}{3}\\
\multicolumn{1}{c|}{\fdc}& \val{0.57}{0.17}     &~~~\val{116}{23} &\val{0.56}{0.17}  &~~~\val{116}{24}      &\val{0.55}{0.18}  &~~~\val{117}{25}     & \val{0.53}{0.18} &~~~\val{117}{26}\\
\multicolumn{1}{c|}{\fnr}& \val{0.59}{0.22}     &~~~\val{137}{14} &\val{0.59}{0.22}  &~~~\val{137}{14}      &\val{0.60}{0.22}  &~~~\val{137}{14}     & \val{0.60}{0.23} &~~~\val{136}{15}\\
\multicolumn{1}{c|}{\fqv}& \val{4.13}{0.09}     &      0       &\val{4.41}{0.10}  &   0               &\val{5.06}{0.11}  &  0               & \val{5.41}{0.11} &   0\\
\multicolumn{1}{c|}{\fds}& \val{1.09}{0.17}     &~\val{-43}{13} &\val{1.14}{0.18}  &~\val{-60}{14}      &\val{2.08}{0.18}  &\val{142}{7}      & \val{2.18}{0.19} &~~\val{125}{5}\\
\multicolumn{1}{c|}{\fvd}& \val{0.81}{0.15}     &~~~~\val{44}{22}  &\val{2.47}{0.17}  &\val{-121}{9}      &\val{0.92}{0.17}  &~~~\val{92}{24}      & \val{2.59}{0.17} &~~\val{-68}{7}\\
\bottomrule
\end{tabular}
}
\end{table}

\cleardoublepage


\addcontentsline{toc}{section}{References}
\bibliographystyle{LHCb}
\bibliography{bib/main,bib/LHCb-PAPER,bib/LHCb-CONF,bib/LHCb-DP,bib/LHCb-ANA,bib/standard.bib,bib/LHCb-TDR.bib}

\newpage
\centerline
{\large\bf LHCb collaboration}
\begin
{flushleft}
\small
R.~Aaij$^{36}$\lhcborcid{0000-0003-0533-1952},
A.S.W.~Abdelmotteleb$^{55}$\lhcborcid{0000-0001-7905-0542},
C.~Abellan~Beteta$^{49}$,
F.~Abudin{\'e}n$^{55}$\lhcborcid{0000-0002-6737-3528},
T.~Ackernley$^{59}$\lhcborcid{0000-0002-5951-3498},
A. A. ~Adefisoye$^{67}$\lhcborcid{0000-0003-2448-1550},
B.~Adeva$^{45}$\lhcborcid{0000-0001-9756-3712},
M.~Adinolfi$^{53}$\lhcborcid{0000-0002-1326-1264},
P.~Adlarson$^{79}$\lhcborcid{0000-0001-6280-3851},
C.~Agapopoulou$^{47}$\lhcborcid{0000-0002-2368-0147}
% ...

{\footnotesize \it

% ...
$^{36}$Nikhef National Institute for Subatomic Physics, Amsterdam, Netherlands\\
% ...
$^{45}$Instituto Galego de F{\'\i}sica de Altas Enerx{\'\i}as (IGFAE), Universidade de Santiago de Compostela, Santiago de Compostela, Spain\\
% ...
$^{47}$European Organization for Nuclear Research (CERN), Geneva, Switzerland\\
% ...
$^{49}$Physik-Institut, Universit{\"a}t Z{\"u}rich, Z{\"u}rich, Switzerland\\
% ...
$^{53}$H.H. Wills Physics Laboratory, University of Bristol, Bristol, United Kingdom\\
% ...
$^{55}$Department of Physics, University of Warwick, Coventry, United Kingdom\\
% ...
$^{58}$School of Physics and Astronomy, University of Glasgow, Glasgow, United Kingdom\\
$^{59}$Oliver Lodge Laboratory, University of Liverpool, Liverpool, United Kingdom\\
% ...
$^{67}$Syracuse University, Syracuse, NY, United States\\
% ...
$^{79}$Department of Physics and Astronomy, Uppsala University, Uppsala, Sweden, associated to $^{58}$\\
% ...
}
\end{flushleft}

\end{document}